\documentclass[letterpaper,twocolumn,10pt]{article}
\usepackage{usenix}

\usepackage{tikz}
\usepackage{amsmath}

\usepackage{soul}

\usepackage{url}
\usepackage{tikz}
\usepackage{amsmath}

\usepackage{xspace}
\usepackage[x11names, table]{xcolor}
\usepackage{booktabs}
\usepackage{float}
\usepackage{graphicx}
\usepackage{makecell}
\usepackage{enumitem}
\usepackage{caption}
\usepackage{multirow}
\usepackage{csquotes}
\usepackage{amssymb} 

\SetBlockEnvironment{displayquote}
  {\vspace{0em}
   \list{}{\leftmargin=1em \rightmargin=1em}%
   \item\relax}%
  {\endlist
   \vspace{0em}}

\newenvironment{squishitemize}
{\begin{list}{\textbullet}{%
    \setlength{\itemsep}{0pt}%
    \setlength{\parsep}{0pt}%
    \setlength{\topsep}{0pt}%
    \setlength{\parskip}{0pt} %
    \setlength{\labelwidth}{.5in}%
    \setlength{\labelsep}{0.05in} %
    \setlength{\leftmargin}{.2in} %
    }}
  {\end{list}}

\usepackage[most]{tcolorbox}

\tcbset{
  beamerblockstyle/.style={
    colframe=Apricot,
    colback=white,
    coltitle=black,
    fonttitle=\bfseries,
    boxrule=0.5pt,
    arc=4pt,
    left=2pt,
    right=2pt,
    top=2pt,
    bottom=2pt,
  }
}

\usepackage[dvipsnames]{xcolor}
\usepackage{tcolorbox}
\usepackage{lipsum}
\tcbuselibrary{skins,breakable}

    {\endtcolorbox}

    {\endtcolorbox}

    {\endtcolorbox}

\usepackage[strict]{changepage}

\usepackage{framed}

\definecolor{formalshade}{rgb}{0.95,0.95,1}

\makeatletter
\tcbset{
    myhbox/.style 2 args={%
        enhanced, 
        breakable,
        colback=white,
        colframe=blue!30!black,
        attach boxed title to top left={yshift*=-\tcboxedtitleheight}, 
        title={#2},
        boxed title size=title,
        boxed title style={%
            sharp corners, 
            rounded corners=northwest, 
            colback=tcbcolframe, 
            boxrule=0pt,
            boxsep=1pt,
        },
        underlay boxed title={%
            \path[fill=tcbcolframe] (title.south west)--(title.south east) 
                to[out=0, in=180] ([xshift=2mm]title.east)--
                (title.center-|frame.east)
                [rounded corners=\kvtcb@arc] |- 
                (frame.north) -| cycle; 
        },
    boxsep=2pt,   
    left=2pt,     
    right=2pt,    
    top=2pt,      
    bottom=2pt,   
        #1
    },
    myvbox/.style 2 args={%
        enhanced, 
        colback=white,
        colframe=blue!30!black,
        left=8mm,
        overlay={
            \node[rotate=90, anchor=north west, inner sep=2mm, text=white] (title@aux) at (frame.south west) {#2};
            \path[fill=tcbcolframe] (title@aux.south west)--(title@aux.south east) 
                to[out=90, in=270] ([yshift=2mm]title@aux.east)--
                (title@aux.center|-frame.north)
                [rounded corners=\kvtcb@arc] -| 
                (frame.west) |- (title@aux.west)[sharp corners] -- cycle;   
            \node[rotate=90, inner sep=2mm, text=white] at (title@aux) {#2};
        },
        #1
    },  
}   
\makeatother

\newtcolorbox{myhbox}[2][]{%
    myhbox={#1}{#2}
}

\newtcolorbox{myvbox}[2][]{%
    myvbox={#1}{#2}
}

\usepackage{ulem}

\newcommand{\workname}{\textsc{Conductify}\xspace}

\usepackage[T1]{fontenc}

\usepackage{booktabs}
\usepackage{tabularx}
\usepackage{array}
\usepackage{multirow}

\definecolor{SpecHeader}{HTML}{D9EAF7}     
\definecolor{SpecSubHeader}{HTML}{E9EEF3}  
\definecolor{SpecLight}{HTML}{F6F8FA}      

\begin{document}
%

\date{}


\title{Analyzing Codes of Conduct for Online Safety in Video Games at Scale}

\author{
{\rm Jiuming Jiang}\\
University of Edinburgh
\and
{\rm Shidong Pan}\\
New York University\\Columbia University
\and
{\rm Daniel W. Woods}\\
University of Edinburgh
\and
{\rm Jingjie Li}\\
University of Edinburgh
} 



\maketitle

\begin{abstract}
Online video games have become major online social spaces where users interact, compete, and create together.
These spaces, however, expose users to a wide spectrum of online harms, including harassment, discrimination, inappropriate content, privacy breach, cheating, and more. 
The shape and severity of such harms vary across game design, mechanics, and community context.
To mitigate these harms, game companies issue Codes of Conduct (CoCs) that articulate online safety rules and direct players to safety resources.
However, it remains unclear how prevalent CoCs are, what  safety, security and privacy violations they govern, and whether they meet growing regulatory and industry expectations.
We develop and leverage \workname, a pipeline for identifying and analyzing CoCs at scale.
Applied to Steam, the largest PC game marketplace, it located the available CoCs for 350 of the 9,586 multiplayer titles on Steam. 
We found that CoCs are more available among popular, adult-oriented, and community-driven games, while most multiplayer games operate without CoCs despite regulatory and industry recommendations. 
Although over 80\% of the games with CoCs available consistently address traditional security and safety violations, their governance approaches vary substantially across types of violations.
A further asymmetry emerges in specificity. Compared with harms related to gameplay mechanics, the articulations of interpersonal harm and the underage player safety are often less specific, despite their relevance to many game communities. 
Together, these results inform the improvement of online safety governance and CoC enforcement practices, and building better safety infrastructure for the community of players and developers.

\end{abstract}


%



\section{Introduction}
Video games have emerged as a key medium for online entertainment, attracting a vast global user base across different ages and cultures~\cite{greitemeyer2022dark, goh2023unravelling}.
Despite the growing popularity, online safety issues on video game platforms remains a pressing concern for service providers, regulators, and players~\cite{kou2021punishment, frey2022governing, ma2023transparency,frommel2024building}.
Players are frequently exposed to a range of online risks, including game exploits that compromise account security and software integrity~\cite{burgess2025inside}, doxxing that causes privacy harms~\cite{Malen2025doxxing}, and harassment targeting minors online~\cite{beres2021don, kilmer2024addressing, schulenberg2023creepy}.
Securing gaming represents a unique challenge because harms often impact at-risk users and arise out of the interaction between gameplay and user interactions~\cite{kou2021punishment}.
For example, aggression between players might be expected on a shooter game, but not a pre-teen farm management game.
In theory, age-rating allows parents to control risks to minors, but in practice these ratings do not capture safety issues emerging from player-to-player interactions~\cite{ChinaGameAge, PEGI, ESRB}.
As a result, online games depend on human and automated moderation to detect and sanction misconduct~\cite{kou2021punishment}. 

To tailor harm prevention measures to each game, the industry has developed a novel form of security policy that describes not only the gaming vendor's practices, but also addresses the behavior of users:
A \textit{Codes of Conduct (CoC)} outlines the behaviors that violate community expectations and provide players with safety resources to encourage responsible play~\cite{busch2015toxic, grace-chiplay22}.
CoCs combine ``soft'' policy to guide players away from harmful behavior with ``hard'' enforcement mechanisms like blocking contents and banning users.
Studying CoCs offers valuable insights into an area of security where designers must operate with an expanded threat model.

Research thus far has consisted of small-scale studies that identify common structural elements and communication approaches in game CoCs through qualitative analysis~\cite{grace-chiplay22}.
However, it remains unclear how widely CoCs have been adopted, and to what extent existing CoCs meet the expectations proposed by the video game industry, policy-makers, and recent online safety regulations, including United Nations Children's Fund (UNICEF)~\cite{UNICEF}, EU Digital Services Act (DSA) 2022~\cite{EU_DSA}, and UK Online Safety Act (OSA) 2023~\cite{UKOnlineSafetyAct}.
To bridge this gap, we present the first large-scale automated analysis to measure and understand CoCs available for online games published on Steam, the largest PC game distribution platform worldwide~\cite{steam}.
Our research addresses three main questions below, motivated by the emerging regulatory landscape for online safety in games:

\begin{itemize}

\item[\textbf{RQ1.}] How does CoC \textit{availability} vary across games?
\item[\textbf{RQ2.}] 
How do CoCs' approaches vary in \textit{governing}
different online safety, security, and privacy violations?

\item[\textbf{RQ3.}] 
How \textit{specific} are CoCs to their gaming and safety contexts? 

\end{itemize}

\textbf{Contributions.} We developed \workname, the first system to automatically analyze video game CoCs. 
First, \workname traverses gaming platforms and surfaces CoCs ``buried'' beyond the homepage (much harder to locate than privacy policies). 
Second, we built a reference dataset of 926 CoC segments annotated with 17 safety-related topics specific to video games, and we trained a classifier that achieves a high overall F1 score of 88.0\% when labeling new CoC documents. 
\workname then extract fine-grained entities from CoCs that characterize a game's interaction and safety contexts, e.g., identified vulnerabilities, for quantitatively and qualitatively evaluate the specificity of a game's CoC in-depth.
Using \workname, we found:
\begin{squishitemize}
\item[\textbf{RQ1.}] 
Games with CoCs tend to be higher-profile, receiving 10 times more player reviews than the average title. CoC availability is also higher among community-driven \textit{Massively Multiplayer Online Role-Playing Games} (16.4\%). Adult-rated titles (\textit{18+}, 7.7\%) and games with higher levels of player-perceived toxicity are also more likely to publish CoCs. However, given the large number of games on the marketplace, most online multiplayer titles (96\%) still lack CoCs, including many child-appropriate and action-intensive games, to respond to industry and regulatory recommendations.

\item[\textbf{RQ2.}] 
The vast majority of games (>80\%) consistently outlaw traditional security violations (cheating via technical exploits), interpersonal safety issues like harassment and discrimination, as well as inappropriate content.
However, the governance approaches against different violations remain inconsistent. 
Games tend to emphasize the punitive consequence rather than tailored moderation mechanisms for violations such as unauthorized transactions that are technically challenging to enforce within the game.

\item[\textbf{RQ3}]
Despite recognizing broad categories of violations similarly, CoCs vary in how specifically they reflect each game’s interaction and safety model. Adult-rated games show higher CoC availability, but their CoCs are not proportionally more distinct. In contrast, community-driven \textit{Massively Multiplayer Online Role-Playing Games} contain more context-specific rules, aligning more closely with industry and regulatory recommendations.
This specificity, however, remains uneven. Although CoCs commonly address interpersonal harms, their treatment of protected demographics is often superficial, as compared to gameplay-related violations and moderation. Beyond listing protected characteristics, only 24\% of titles mention child-related safety issues, and these mentions do not align with their targeted age groups.

\end{squishitemize}

Our insights further provoke critical directions to improve online safety governance, including alternative models of CoC adoption, mediating tensions between developers and players with better transparency, and improving integration of CoC and safety resources for a large number of developer and player communities.

\section{Background and Related Work}

\noindent
\textbf{Online safety in video games.}
Online safety in video games represents a persistent and evolving challenge, one that has drawn increasing attention from researchers over the past two decades~\cite{kiene2019volunteer, kou2021punishment, frey2022governing, ma2023transparency}.
These issues span a wide range of disruptive behaviors that violate social norms, community rules, and user expectations around security and privacy~\cite{beres2021don}. Such risks not only create hostile environments that diminish player experiences~\cite{canossa2021for}, but also cause tangible harms including emotional distress~\cite{reid2022feeling}, privacy breaches~\cite{martinovic2014you},  and marginalization in gaming communities~\cite{seng2019analyzing}.
Harassment and hate are particularly salient concerns. Beyond verbal abuse via in-game communication~\cite{zorah2020i, kou2021punishment}, toxic players and online predators often exploit game mechanics and interfaces~\cite{zhang2024toxicity}—for example, through stalking, offensive gestures, or threatening use of avatars~\cite{frey2022governing}. At the technical level, bugs and software vulnerabilities enable cheating and the theft of copyrighted or privacy-sensitive information, posing ongoing threats despite industry countermeasures~\cite{burgess2025inside}.
These safety issues are deeply intertwined with the very qualities that make games engaging--interactivity, competitiveness, and immersion--making them difficult to address. For example, the high-stakes, competitive design of Multiplayer Online Battle Arena (MOBA) games can exacerbate conflict and hostility among players, fueled by perceived loss, powerlessness, and the contagion of toxic behaviors~\cite{kou2020toxic, kordyaka2020towards, canossa2021for}. Similarly, monetization mechanisms like loot boxes incentivize cheating and exploitation, including account boosting and unauthorized real-money trading~\cite{burgess2025inside}. Unlike social media platforms, where safety concerns primarily involve harmful content and verbal abuse, the complexity of gameplay mechanics creates a unique and multifaceted safety landscape.

Game communities also draw highly diverse audiences, with a substantial share of minors. In the UK, 93\% of children report playing video games~\cite{UK_Ofcom}. Yet minors are especially vulnerable due to limited digital literacy and developmental capacity to recognize and respond to risks~\cite{song2025predatory, zhang2025dangerous}. They may also be mislabeled as ``griefers'' not out of malice, but due to unfamiliarity with unspoken community norms~\cite{lin2005the, kordyaka2023the}, leading to stigmatization and exclusion from gaming communities.
Safety risks extend beyond age. Discrimination and hate based on gender, sexual orientation, disability, or cultural background remain pervasive~\cite{frey2022governing, hernandez2025who}. Gender-based harassment is especially widespread: a recent study found that 56.6\% of female players reported experiencing online harassment~\cite{femalegamereport}. 
While women’s participation in esports has grown, entrenched masculine cultures continue to fuel bias and harassment~\cite{tang-hssc25}. In many communities, such harms are normalized~\cite{beres2021don}, with repeated exposure to abusive content reinforcing desensitization and aggressive behavior even beyond the game context~\cite{lin2005the}.

\noindent
\textbf{Safety governance and CoCs.}
Video game companies and regulators have developed governance approaches across different stages of game distribution and play.
One common mechanism is game rating systems, including the Pan-European Game Information (PEGI)~\cite{PEGI} and Entertainment Software Rating Board (ESRB) in North America~\cite{ESRB}.
These systems classify games by content (e.g., violence, horror) and assess age suitability. Over time, ratings have become legal requirements in several countries such as Germany, which makes their age rating USK mandatory for game publishing~\cite{xiao-gem24}. 
Game distribution platforms also display these ratings to guide consumers, particularly parents.
Despite their utility in assessing content safety pre-release, age ratings fall short in addressing player interactions~\cite{xiao-rs23, frey2022governing}.

Another line of prior work has focused on moderation strategies against unsafe in-game behavior, including improving victim–moderator communication~\cite{kou-cscw21}, integrating machine learning with human oversight~\cite{kou-chi21}, and designing safeguards for younger players~\cite{tekinbacs-tochi21}. Yet, many games still lack usable moderation mechanisms~\cite{faraz2022child}.
Barriers persist: moderation is time-consuming and error-prone for both humans and automated systems~\cite{sabri-chi23}; players often find ways to bypass restrictions (e.g., nicknames)~\cite{boustani-chiplay20}; privacy concerns arise from the information needed in automated moderation~\cite{dolling-ai25}; and lack of transparency results in the perceptions of unfair moderation~\cite{ma2023transparency}.
The complexity and variation of player behaviors make game moderation a distinct challenge compared to social media where text content moderation is primary~\cite{li-ndss24}.

In response, researchers and practitioners have increasingly advocated for the standardization of \textit{Codes of Conduct (CoCs)} as a proactive governance tool that ``nudges players towards better behavior''~\cite{busch2015toxic}.
Beyond gaming, CoCs have been widely adopted in education\cite{rezaee-jbe01}, business~\cite{adam-jbe04}, and software engineering~\cite{gotterbarn-acmcomm97}. 
In video games, CoCs describe unacceptable behaviors and associated safety, security, and privacy risks, codifying specific community expectations and norms in ways that go beyond reactive moderation~\cite{grace-chiplay22}.

\noindent
\textbf{Governance document analysis for online services.}
Online service providers are responsible for implementing measures that safeguard user security, privacy, and safety.
Governance information is typically communicated through online documents, such as privacy policies, which are mandated by regulations and intended to provide transparency. 
Their accessibility has enabled large-scale empirical studies and the development of automated tools for collection and analysis~\cite{harkous2018polisis, del2022systematic, lin2024automated, pan2024trap, pan2024hope}.
Nevertheless, availability remains inconsistent: only 37\% of Alexa Top–10k websites provide a privacy policy~\cite{amos2021privacy}, and 15.7\% of apps lack a policy link on their homepage~\cite{pan2024trap}. 
Similar gaps are observed for Terms of Service adoption~\cite{sundareswara2021large}.

The process in automated governance document analysis is driven by advances in natural language processing (NLP) models and relevant datasets, including OPP-115 annotated for privacy policies~\cite{wilson2016creation}.
For privacy policies, researchers have explored several NLP paradigms.
Rule-based approaches rely on curated lexicons, regular expressions, and semantic templates to identify normative statements about user rights and obligations~\cite{oltramari2018privonto, andow2020actions, cui2023poligraph}; Supervised and unsupervised machine learning methods classify, extract, and summarize policy content, adapting to linguistic diversity~\cite{liu2018towards, pan2024hope}. More recently, transformer-based models, including domain-adapted \texttt{BERT} variants, and generative models such as \texttt{GPT} have enabled more fine-grained and context-aware analyses of governance documents~\cite{pan2023toward, chen2025clear, shanmugarasa2025privacy, xie-sec25}.

In contrast, safety governance documents, including CoCs, remain underexplored due to both the lack of standardized, scalable analysis frameworks and the difficulty of locating these documents.
Most existing work is qualitative in nature and focuses on content moderation policies. 
Studies on major social media platforms (e.g., Facebook and Reddit) highlight important governance practices~\cite{chandrasekharan2018internet, fiesler2018reddit, kuo2023unsung, gao-sec25}, while they argue that findings from large platforms may not generalize to smaller or more specialized communities~\cite{singhal2023sok}.
Video games present a distinct and understudied context where online safety issues are shaped by gameplay mechanics, technical infrastructures, and player dynamics. Prior research has qualitatively examined video game CoCs, identifying common structural elements and presentation strategies by manually coding a small sample of 32 CoCs~\cite{grace-chiplay22, busch-vgp15}.
However, the extent to which CoC documents have been adopted by commercial games and game environments remains unknown.

\section{Research Question Formulation}
\label{sec_roadmap}
Unlike privacy governance, where frameworks such as the GDPR and privacy policy requirements are well established, online safety governance in video games remains emerging and underexplored. Recent regulations, including the EU DSA 2022~\cite{EU_DSA} and UK OSA 2023~\cite{UKOnlineSafetyAct}, increasingly emphasize platforms' responsibilities to proactively protect users. This motivates our study of CoC adoption as a key mechanism for proactive safety governance in games.
Our research questions are grounded in expectations for CoCs derived across online safety regulation, policy, and industry guidance, including UNICEF's recommendations for the gaming industry~\cite{unicef-recommend26}, the Fair Play Alliance's online harm framework~\cite{FPA_Framework, thrivingingames-harm26}, the EU DSA (\S14 and \S17), and UK Office of Communications (Ofcom)'s Protection of Children Code of Practice~\cite{ofcom-protecting26}.

\begin{itemize}
    \item \textbf{CoC availability (RQ1)}. 
    Game services should provide accessible CoCs that clearly communicate prohibited behaviors and safety resources. We examine how widely CoCs are available, which types of games lead or lag in adoption, and whether games serving underage players provide adequate access to such guidance. 

    \item \textbf{Violation governance (RQ2).} 
    CoCs should identify violations and enforcement procedures that are relevant to the service and its safety, security, and privacy risks. We examine how CoCs in govern various safety, security, and privacy violations through moderation measures and restrictions.

    \item \textbf{Context specificity (RQ3).} 
    CoCs should reflect the specific game environment, including gameplay mechanics, interaction features, target communities, and moderation procedures. We examine how far published CoCs are adapted to different gaming contexts.
\end{itemize}

\begin{figure*}[h!]
    \centering
       \includegraphics[width=\textwidth]{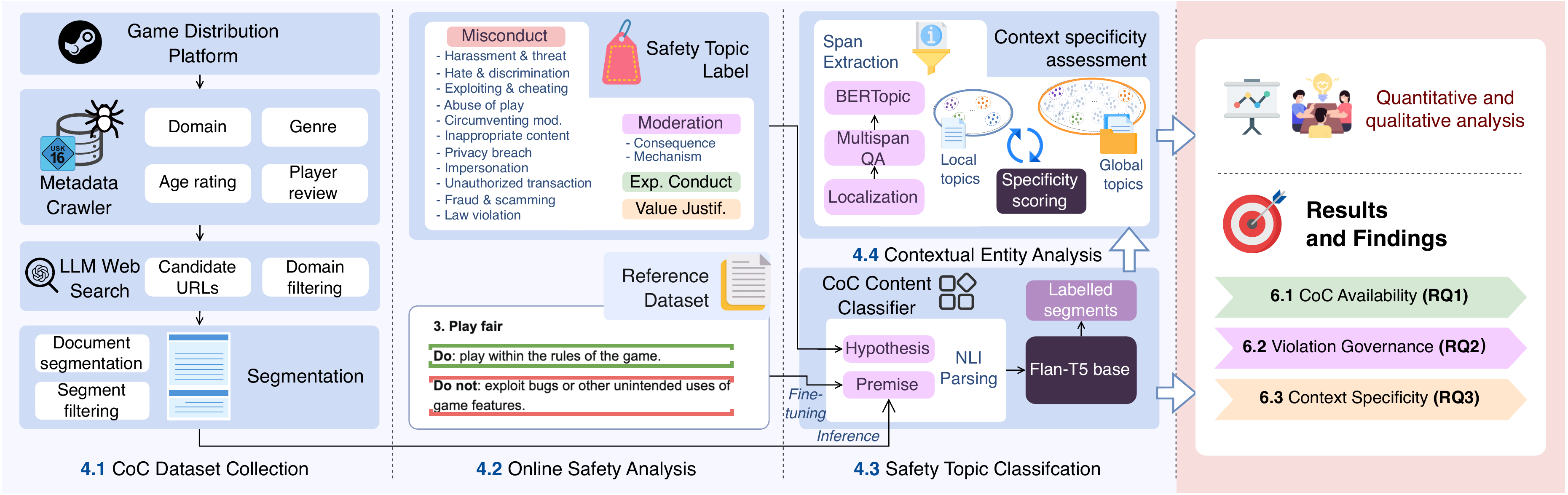}
    \caption{\workname framework overview.} 
    \label{fig:Conductify framework overview}
\end{figure*}

\section{Methodology}
\label{sec_methodology}
Systematically addressing above questions remained challenging due to the lack of a unified interface for presenting CoCs on game platforms and frameworks and resources for automatically extracting and analysing safety-relevant content in CoCs.
We tackle these challenges by building \workname, an end-to-end NLP-powered pipeline. To address \textbf{RQ1}, \workname identifies CoCs at scale and analyzes their availability across games identified from Steam, the world's largest PC game marketplace, that vary in targeted age groups, player-perceived toxicity and popularity, and genres (Section~\ref{sec_Results_Adoption}); 
for \textbf{RQ2}, \workname structurally segmented CoCs and classifies 11 specific forms of misconduct identified as violations and how they are moderated, then cross-compares different types of games for quantitative analysis (Section~\ref{sec_Results_Misconduct}); furthermore, \workname measures how specific and distinctive CoCs are by extracting fine-grained elements that describe the online safety contexts in games (Section~\ref{sec_results_Moderation}), and we further present our qualitative insights for \textbf{RQ3}. 
Figure~\ref{fig:Conductify framework overview} provides an overview of our study pipeline and \workname's core modules, described in detail below.

\subsection{CoC Dataset Collection} 
\label{sec_Method_Dataset}
\noindent
\textbf{Video game metadata crawler.} 
Video game distribution platforms act as marketplaces that connect developers and players, providing information that guides purchasing decisions thus offering a useful proxy for studying the real-world game ecosystem. 
This work focuses on Steam~\cite{steam}, the world’s largest PC gaming marketplace, which hosts over 147 million monthly active users as of  2025~\cite{steam_stat_2025}.

We identify most relevant attributes and associated data of Steam games for identifying CoCs and contextualizing our analysis. 
For each \textsf{Game title}, we collect the following in Table~\ref{tab:game-metadata}.
To collect this information, \workname employs a \texttt{Playwright}-based crawler to build a comprehensive index of Steam games, and then retrieves additional attributes and data using the official Steam Web API~\cite{steamapi}.

\begin{table}[t]
\footnotesize
\centering
\caption{Associated game data collected for CoC analysis.}
\label{tab:game-metadata}
\setlength{\tabcolsep}{4pt}
\renewcommand{\arraystretch}{1.08}

\begin{tabularx}{\columnwidth}{
    >{\raggedright\arraybackslash}p{0.25\columnwidth}
    X
}
\toprule

\rowcolor{SpecSubHeader}
\textbf{Field} & \textbf{Description} \\
\midrule

\rowcolor{SpecSubHeader}
\textsf{Domain name}
& Official domain names of the game and its developer. \\

\midrule
\rowcolor{SpecSubHeader}
\textsf{Age rating}
& Minimum player age threshold based on pre-release content assessment. We use Steam's USK-aligned rating mandated in Germany~\cite{steam-germany25}, as it provides broader coverage than optional systems such as PEGI~\cite{PEGI} or ESRB~\cite{ESRB}. \\

\midrule
\rowcolor{SpecSubHeader}
\textsf{Game genre}
& High-level categorization of game design that shapes player interactions and potentially the safety context. Following prior work~\cite{10.1145/3772318.3791085}, we focus on: \\

\rowcolor{SpecLight}
\quad \textit{MOBA}
& Multiplayer Online Battle Arena (MOBA) features highly competitive gameplay with temporary teams, matching and ranking. \\

\rowcolor{SpecLight}
\quad \textit{Battle Royale}
& Battle Royale rewards last player/team-standing in anonymous large-scale competitions with ranking. \\

\rowcolor{SpecLight}
\quad \textit{Shooter}
& Shooter games involve solo or team-based competition using ranged weapons on instanced maps. \\

\rowcolor{SpecLight}
\quad \textit{MMORPG}
& Massively Multiplayer Online Role-Playing Game (MMORPG) highlights large persistent world and narratives, involving player groups with long-term interdependency and reputation for quests, cooperation, and competition.   \\

\rowcolor{SpecLight}
\quad \textit{Sports}
& Sports games involves head-to-head rivalry with realistic arenas and short seasonal cycles.   \\

\rowcolor{SpecLight}
\quad \textit{Fighting}
& Fighting games are tournament-driven competitions in instanced arenas.   \\

\rowcolor{SpecLight}
\quad \textit{Sandbox}
& Sandbox games include persistent but fragmented communities in open world that features cooperation and creativity.  \\

\midrule
\rowcolor{SpecSubHeader}
\textsf{Player review}
& We collected player reviews on Steam comments from verified purchasers as a proxy to measure \textit{popularity} and \textit{player-perceived toxicity}. \\

\bottomrule
\end{tabularx}
\end{table}

\noindent
\textbf{Game title filtering.}
To remove duplication and retain only in-scope video game titles, \workname filters games based on collected metadata. 
First, we exclude entries of Downloadable Content (DLC), as these are just extensions of base games.
Second, our work focuses on \textit{Online Multiplayer} titles, which are most relevant to online safety issues arising from player interactions.
To identify such games, we leverage Steam’s three multiplayer play-mode tags: \textit{MMO}, \textit{Online PvP}, and \textit{Online Co-op}.

\noindent
\textbf{LLM-powered CoC retrieval.}
Unlike privacy policies, which are typically placed in website headers or footers, CoCs lack standardized presentation and are therefore harder to locate.
To address this, \workname leverages a state-of-the-art web search API based on OpenAI’s \texttt{GPT-4o-search} model~\cite{gpt4osearch}, which translates natural-language prompts into queries for the partnered search engine and returns structured responses. 
We carefully design and test our search prompt to maximize retrieval of relevant CoC documents.
For each game, \workname uses \texttt{gpt-4o-search} to obtain a list of candidate URLs and identify whether a page contains a \textit{standalone} CoC—i.e., a document whose primary purpose is to guide players on behavioral safety.
It then filters these URLs against the game’s official domains and verifies their validity via \texttt{HTTP} responses. 
Two researchers independently verified the performance of this module on 100 randomly sampled games, which shows a satisfactory accuracy of 91\% in correctly identifying the standalone CoCs associated with the game.
We provide our prompts in our artifact.

\noindent
\textbf{CoC segmentation.} 
For each confirmed CoC URL, \workname downloads the corresponding \texttt{HTML} file and segments it into chunks for fine-grained classification.
The raw \texttt{HTML} is sanitized and reformatted by removing stylistic elements while preserving essential structural tags such as headings, paragraphs, and lists.
We then apply a machine learning–based tool~\cite{pdfanalysis-github25} as the segmenter, which ingests the CoC document converted in \texttt{PDF} format and extract text blocks using a pretrained model.
Additionally, we retain only English-language segments for subsequent analysis.

\subsection{Systemizing Online Safety Analysis in CoC}
\label{sec_framework_safetyanalisis}

Despite many prior works focusing on privacy document analysis by NLP~\cite{wilson2016creation, harkous2018polisis, pan2023toward, cui2023poligraph}, how to systematically analyze online safety CoC documents of video games remains understudied. 
Thus, we build a representative analysis framework and reference dataset for developing NLP modules to analyze CoC documents automatically.  

\noindent
\textbf{CoC segment topic.}
We start from a deductive approach to identify the main topics in CoC documents. 
Through reviewing key literature and industry standards for CoC design~\cite{grace-chiplay22}, moderation policies~\cite{singhal2023sok, schaffner2024community},  and online safety in gaming~\cite{thrivingingames-framework}, we identify four high-level topics that are common across CoC segments:

\begin{squishitemize}
\item  \textsf{Misconduct}: -- behaviors that are considered as violation and do not align with the norms and rules of the game, its community and service provider. 

\item  \textsf{Moderation} -- the punishment applied if a rule of the game, its community and service provider is violated as well as the procedure to report and detect such violations.

\item \textsf{Expected conduct} -- behaviors that are expected and encouraged for players to adhere with the norms and rules of the game, its community and service provider. 

\item  \textsf{Value justification} -- the values and core beliefs of the game, its community and service provider that are used to justify a CoC term in the above categories.

\end{squishitemize}

\textsf{Misconduct} and \textsf{Moderation} are most relevant to our research focus, as they describe the \textit{specific online safety violations} considered by a game and \textit{recourse to support players' safety needs}. 
\textsf{Expected conduct} and \textsf{Value justification} additionally indicate how games positively \textit{communicate} users towards safer behaviors and cultures.  

Therefore, we dive into \textsf{Misconduct} and \textsf{Moderation}, identifying 13 key subtopics.
These include 11 specific forms of misconduct: (1) abusive inter-personal behaviors of \textsf{Harassment and threat} and \textsf{Hate and discrimination}; (2)  \textsf{Exploiting and cheating}, \textsf{Abuse of play}, and \textsf{Circumventing moderation} that exploit or compromise game mechanics; (3)  informational risks of \textsf{Inappropriate content} and \textsf{Privacy breach}, as well as \textsf{Impersonation} that threatening player identities; (4)  additionally, \textsf{Unauthorized transaction}, \textsf{Fraud and scamming}, and \textsf{Law violation} that may incur legal risks.

We initially identified misconduct types from established game safety frameworks~\cite{grace-chiplay22, thrivingingames-framework}, which grouped misconduct based on shared concepts (\textsf{Hate and discrimination} concerns identity-based harm, while \textsf{Harassment and threat} highlights behavior patterns).
Through our further data exploration and annotation described next, we recognized, merged and refined low frequency and highly overlapping concepts, e.g., ``stalking'' as a form of harassment.

For \textsf{Moderation}, we consider two subtopics if a CoC segment mentions specific \textsf{Moderation consequence} or \textsf{Moderation mechanism}.
.
Full subtopic definitions are available in Appendix~\ref{sec_appendix-label}.

\noindent
\textbf{Reference dataset.}
To develop NLP models for \workname to classify and analyze online safety–related topics, we construct and annotate a reference dataset including 67 representative CoC documents. 
Our sampling draws on the 42 video game companies, spanning diverse sizes and game genres, based on those identified by Grace et al.'s qualitative analysis~\cite{grace-chiplay22}. 
CoCs were manually collected from their official websites between June and August 2024.

We segment the collected documents and annotate each segment using the label scheme described above. 
Each segment may receive multiple labels.
Our data exploration reveals a skewed distribution, with \textsf{Misconduct} dominating other categories. 
To mitigate imbalance for efficient training, we apply a zero-shot classification with \texttt{GPT-4o} to sample segments across 67 documents, targeting a more even distribution of the four high-level topics.
A random subset of 11 documents (including 334 segments) is held out in full in our annotation as the test set for model evaluation; our training set includes 592 segments from 56 documents.
Our manual annotations did not rely on \texttt{GPT-4o}’s suggested topics.

Our team is made up of domain experts in computer security and privacy, cyber policy, human-computer interaction, and game development. 
Two primary authors first performed independent annotations then finalize the dataset with other team members:
After completing the initial annotation, all team members hold meetings to discuss notes taken on label criteria and ambiguities, which are later discussed and resolved in team meetings.
Then both annotators re-review the annotations and correct the labels independently with a third team member verify and resolve outstanding disagreements.
Our overall inter-rater reliability (Cohen’s Kappa) before curation was $\kappa = 0.725$, indicating substantial level of agreement. 
The breakdown is available in Table~\ref{tab:label_stats} of Appendix~\ref{sec_appendix_irr}.

\noindent
\textbf{Contextual entities.}
Our data exploration shows that tailored safety guidelines are expressed through fine-grained entities (e.g., specific terms and phrases) that provide a deeper look into a game's interaction and safety governance models, serving as an indicator to measure the specificity of CoCs.
For example, a tailored CoC will specify relevant community members whom the game aims to protect and which specific features may be exploited in a violation. 
Compared to privacy-sensitive elements, which are well defined such as information types and data-use purposes~\cite{wilson2016creation, bui-pets21}, safety-related entities are articulated in CoCs remains underexplored.
To address the above and measure the context specificity (\textbf{RQ3}) in CoCs, we will extract and analyze such contextual entities from CoCs in two groups that are key elements in the safety and governance model of video games~\cite{FPA_Framework}: \textsf{Misconduct}-related entities (\textsf{target of protection}, \textsf{vulnerabilities and exploits}, and \textsf{inappropriate information}), and \textsf{Moderation}-related entities (\textsf{roles in moderation}, \textsf{consequences}, and \textsf{mechanisms} for addressing misconduct).

\subsection{Online Safety Topic Classification}
We introduce our design of \workname's NLP model to accurately and automatically classify various safety topics and risks discussed in CoC.

\noindent
\textbf{Topic classification through natural language inference.}
\workname's segment classifier assigns topic labels on each CoC segment extracted.
To efficiently train a classifier for 4 high-level and 13 safety-related subtopics, we formulate the multi-label classification problem as a natural language inference (NLI) task~\cite{harkous-sp22, camburu-nips18}. 
Each CoC segment serves as the \textit{premise}, while each label definition is phrased as a \textit{hypothesis} (e.g., ``This text is about Misconduct''). The model predicts whether the premise entails the hypothesis, yielding a binary classification outcome.
Repeating this process across labels allows a single NLI model to classify multiple topics and subtopics. Appendix~\ref{sec_appendix_nli_template} show an example of our NLI template.
For each segment, \workname queries the NLI model with all 17 labels, producing a set of binary predictions. To preserve label hierarchy consistency, we apply post-processing: if a subtopic is predicted as positive while its parent topic is not, the parent topic is automatically marked as positive.

\noindent
\textbf{Model training.}
We use \texttt{FLAN-T5-Base}, an instruction-tuned model well-suited for NLI and data-efficient learning~\cite{flan_t5_base}, as the backbone for high-level topic classification. 
We fine-tune the model on our training set of 592 CoC segments, applying oversampling to mitigate class imbalance and improve representation of minority topics. 
Appendix~\ref{sec_appendix_model_config} discusses training parameters in details.

\noindent
\textbf{Performance evaluation.}
We evaluate our topic classifier using the annotated test set including 334 segments and prove its accuracy for quantitative measurement.
Our model performance for each topic is shown in Table~\ref{tab:rest5l1}.
Consistent with prior work, we report macro-average F1 scores~\cite{harkous2018polisis}. 
The results show that our NLI classification using \texttt{FLAN-T5-Base} achieves a high overall F1 score of 0.88, and it is comparable to related prior work on privacy policy analysis and measurement~\cite{wilson2016creation, harkous2018polisis, liu2021have, pan2023toward}.

\noindent
\textbf{Inference and quantitative analysis.}
We leverage \workname's NLI model to classify CoC documents for Steam games. 
We aggregate inference results and metadata for each game to allow further quantitative analysis of (sub)topic coverage. 
To optimize for precision of our measurement results, \workname further sanitizes list of retrieved documents, e.g., that may include ``soft'' 404 pages and generic landing pages, supported by inference results -- we exclude documents without a specific subtopic.


\begin{table}[t]
    \footnotesize
    \setlength{\tabcolsep}{3.5pt}
    \renewcommand{\arraystretch}{0.9}
    \centering
    \caption{Performance evaluation of \workname's topic classifier (per label, macro average) on 17 labels. }
    \label{tab:rest5l1}
    \begin{tabular}{m{3.8cm} >{\centering\arraybackslash}m{0.7cm} >{\centering\arraybackslash}m{0.7cm} 
                >{\centering\arraybackslash}m{0.7cm} >{\centering\arraybackslash}m{0.7cm} 
                >{\centering\arraybackslash}m{0.7cm}}
    \toprule
        \textbf{Label} & \textbf{Prec.} & \textbf{Recall} & \textbf{F1} & \textbf{Acc.} & \textbf{Support} \\
    \midrule

    \rowcolor{blue!35}\multicolumn{6}{l}{\textbf{Misconduct}} \\
    \rowcolor{blue!10}\quad Misconduct (Topic) & 0.921 & 0.907 & 0.910 & 0.912 & 174 \\
    \quad Harassment and threat & 0.861 & 0.876 & 0.868 & 0.924 & 56 \\
    \quad Hate and discrimination & 0.884 & 0.905 & 0.894 & 0.979 & 17 \\
    \quad Exploiting and cheating & 0.949 & 0.920 & 0.934 & 0.973 & 40 \\
    \quad Abuse of play and antagonistic play & 0.911 & 0.644 & 0.704 & 0.945 & 24 \\
    \quad Circumventing and abusing moderation mechanism & 0.944 & 0.845 & 0.887 & 0.985 & 13 \\
    \quad Inappropriate content creation and sharing & 0.837 & 0.837 & 0.837 & 0.896 & 65 \\
    \quad Privacy breach & 0.828 & 0.901 & 0.860 & 0.970 & 17 \\
    \quad Impersonation and identity theft & 0.926 & 0.830 & 0.871 & 0.973 & 21 \\
    \quad Unauthorized transaction and commercialization & 0.947 & 0.908 & 0.926 & 0.991 & 11 \\
    \quad Fraud and scamming & 0.863 & 0.917 & 0.888 & 0.982 & 13 \\
    \quad Law violation & 0.939 & 0.883 & 0.909 & 0.979 & 22 \\

    \rowcolor{green!60}\multicolumn{6}{l}{\textbf{Moderation}} \\
    \rowcolor{green!20}\quad Moderation (Topic) & 0.891 & 0.927 & 0.907 & 0.927 & 82 \\
    \quad Moderation consequence & 0.928 & 0.912 & 0.920 & 0.963 & 44 \\
    \quad Moderation mechanism   & 0.906 & 0.906 & 0.906 & 0.963 & 36 \\

    \rowcolor{gray!20}\multicolumn{6}{l}{\textbf{Positive governance}} \\
    \quad Expected conduct & 0.875 & 0.887 & 0.881 & 0.893 & 107 \\
    \quad Values justification & 0.842 & 0.880 & 0.858 & 0.896 & 73 \\

    \midrule
    \textbf{Overall (macro)} 
    & 0.897 & 0.876 & 0.880 & 0.950 & - \\
    \bottomrule
    \end{tabular}
\end{table}


\subsection{Contextual Entity Extraction and Analysis}
\label{sec_Method_Key_Entity}
Based on topic-classified segments, \workname further extracts and analyzes fine-grained key entities that capture how CoCs specify the context for safety governance.

\noindent
\textbf{Question answering (QA)-based entity extraction.}
Traditional named entity recognition models are insufficient our task, as they focus on generic entity types (e.g., people, organizations, dates)~\cite{li-kde20} rather than the safety-related entities of interest. 
We address this limitation using an open-domain question-answering approach.
Specifically, \workname adopts the QASE framework with \texttt{FLAN-T5-Large}~\cite{flan_t5_base} in an extractive QA setting, which returns text spans directly from the input and is less prone to hallucination than abstractive methods. This feature makes the approach flexible and reliable for our exploratory analysis.
We operationalize entity extraction by mapping each key entity type to one of six QA queries. Before extraction, \workname narrows the input to relevant CoC segments using predicted topic labels (e.g., \textsf{Misconduct}- or \textsf{Moderation}-related), reducing noise and improving efficiency. The model may return multiple candidate spans per query. 
Appendix~\ref{sec_appendix_qa_template} shows an example of our query, and we provide a complete list of QA queries is provided in our artifact.

To improve answer quality, we sanitize the extracted spans by
filtering incomplete spans by checking against the span boundaries and the \texttt{nltk}~\cite{bird2006nltk} dictionary. 
We verified the performance of our entity extraction by manually validating 120 returned spans across six questions.
Our result shows that \workname achieves an satisfactory accuracy of 75\% when extracting relevant spans.

\noindent
\textbf{Context specificity assessment.} 
We use the extracted entity spans to measure how context-specific each CoC is relative to the full corpus. 
We cluster these spans into topic models, supporting robust comparison, using \texttt{HDBSCAN}~\cite{mcinnes2017hdbscan} in \texttt{BERTopic}~\cite{grootendorst2022bertopic}. To preserve context, \workname embeds each span using token representations from its original segment using  embeddings from \texttt{all-MiniLM-L6-v2}~\cite{all_minilm_l6_v2}.

We cluster extracted span entities with \texttt{HDBSCAN}~\cite{mcinnes2017hdbscan} in \texttt{BERTopic}~\cite{grootendorst2022bertopic} to build topic models for more reliable comparison. 
To preserve context, \workname embeds each span using contextualized token representations from its original segment. 
For each segment $s$, we run a pre-trained  encoder (\texttt{all-MiniLM-L6-v2}~\cite{all_minilm_l6_v2}) once, and mean-pool the token embeddings whose character offsets overlap with span $e$. 
We then fit the \texttt{HDBSCAN} model on the resulting span embeddings.

Following the modified Hausdorff distance~\cite{dubuisson_modified_hausdorff}, we define context specificity as the mean nearest-neighbor distance between a CoC's local topic centroids and the topic centroids of the topic model built on the rest of the corpus with other games' CoCs. 
Let $\mathcal{T}_g = \{\mathbf{c}_1^g, \dots, \mathbf{c}_{K_g}^g\}$ be the centroid embeddings of topics for subject $g$ (e.g., a game), and let $\mathcal{T}_G = \{\mathbf{c}_1^G, \dots, \mathbf{c}_{K_G}^G\}$ be the centroids from the global counterpart topic model. 
Using cosine similarity $\mathrm{sim}(\cdot,\cdot)$, we compute: $
\ell(\mathcal{T}_g, \mathcal{T}_G) 
= 1 - \frac{1}{K_g}
\sum_{\mathbf{c}^g \in \mathcal{T}_g}
\max_{\mathbf{c}^G \in \mathcal{T}_G}
\mathrm{sim}(\mathbf{c}^g, \mathbf{c}^G).
\label{eq:ell1} $

For each local topic, this metric finds its closest global counterpart and averages the remaining distance. 
Higher values indicate that a CoC contains concepts less represented in the wider corpus. 
We use this score to compare specificity across games and across elements of the interaction and safety models of a game.
For each game, we obtain specificity measurement by leave-one-out to build counterpart topic models.

\noindent
\textbf{Content analysis.}
We conducted a qualitative content analysis of contextual entities that characterize the interaction and safety models in CoCs, informed by the topic models and specificity assessment performed above. 
The extracted entities, with resulting topic clusters and specificity scores, are iteratively reviewed within the context of their corresponding segments by the research team collaboratively. 
We further refined the topic clusters through our team discussions by merging overlapping clusters and assigning interpretable labels.
This process transforms raw entity spans into coherent categories that anchor our content analysis and identity themes to characterize what elements make community-governance specific, and how.

\section{Limitations}
Firstly, we focus on \textit{standalone} CoC, as a key expectation to provide accessible communication, published on the \textit{official} game company domains in English. 
While we recognize that player communities may adopt unofficial CoCs and that other corporate documents (e.g., terms of service) may also include behavioral rules, our choice to focus on official standalone CoCs is intentional. These documents most directly reflect how companies formally approach safety governance, whereas the scope of other documents often differs (e.g., service protection rather than player-to-player safety). Future work could expand to these complementary sources for a more comprehensive view of governance.
Second, \workname’s CoC retriever could be improved with a more complete database of game companies. Although Steam offers broad coverage, subsidiaries and unlisted domains can lead to false negatives.
Third, one could explore alternative NLP approaches suitable for different application contexts, such as prompt-engineered LLMs, which could reduce data requirements. Nevertheless, \workname’s open-source contribution is model-agnostic. This work chooses T5 that benefits transparency and local deployment for researchers with limited LLM access.
Furthermore, we measure popularity and player-perceived toxicity by analyzing public reviews on Steam as a proxy estimate, which may be subject to disclosure bias. 
We advocate future work to systematically examine the alignment between player-perceived harms and CoC enforcement in real world contexts. 
Despite these natural boundaries, \workname presents a significant contribution as the first systematic study of online safety CoCs in real-world  games and player communities.

\section{Results and Findings}

Section~\ref{sec_Results_Adoption} to~\ref{sec_results_Moderation} present results from the games and CoCs measured by \workname, answering RQ1, RQ2, and RQ3 about CoC availability, governance for safety violations, and specificity for individual game contexts, respectively.
As of July 31, 2025, \workname identifies 9,586 Steam multiplayer titles.
Many titles even lack basic signals and channels for governance: 3,463 have no official website, and 1,540 have no USK-aligned age rating.
350 games have a retrievable, valid CoC page.
The 350 games map to 261 unique CoC documents, as titles may share a company-wide CoC.


\subsection{RQ1: CoC Availability}
\label{sec_Results_Adoption}
We examine how available are official CoC documents across different categories of online video games.
Our analysis categorizes games by (1) \textit{game genres}, capturing the interaction contexts by design, (2) \textit{age rating}, indicating different levels of content ``harmfulness'' for young audience and adults,  (3) \textit{player-perceived toxicity} that reflect the overall prevalence of harmful behaviors within a game community~\cite{wijkstra_rogers_mandryk_veltkamp_frommel_2024}, and (4) \textit{popularity}, which reflects community sizes.
We observe the following insights.

\noindent
\textbf{CoC availability is concentrated in well-resourced and popular games.} 
We find a significant correlation between the availability of official CoC and their popularity (count of reviews from verified players), using logistic regression ($LR\ \chi^2$ = 618.33, $p < 0.001$, $dof = 1$).
Games with an available CoC are typically high-profile titles, averaging 73,959 Steam reviews, which is approximately 10 times higher than the average review count of multiplayer games (7,364), such as \textit{Rainbow Six}, leaving the vast majority--96.35\% of multiplayer games on Steam--lacking such governance.

\begin{figure}[t]
    \centering
    \includegraphics[width=\columnwidth]{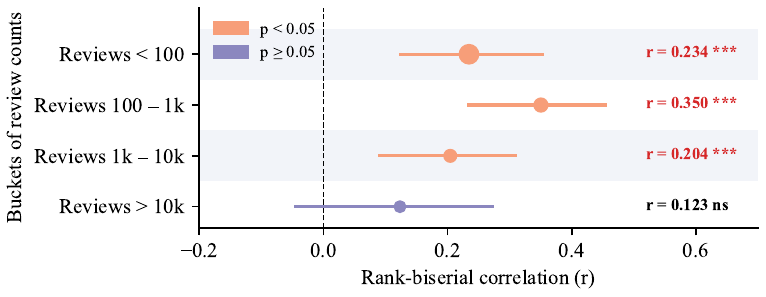}
    \caption{Effect of CoC presence on prevalence of player-perceived toxicity by review-count stratum.} 
    \label{fig:tox_forest}
\end{figure}

\noindent
\textbf{The availability of CoCs is associated with elevated 
levels of player-perceived toxicity and stronger security enforcement.}
To estimate the prevalence of player-perceived toxicity within a game community, we apply a lexicon-based approach to identify player reviews related to toxicity encounters and complaints for each game, adopting the search terms (Appendix~\ref{keyword_list_lexicon}) developed in prior work~\cite{wijkstra_rogers_mandryk_veltkamp_frommel_2024}. 
We measure the toxicity prevalence by calculating the percentage of reviews complaining about toxicity  among the total number of reviews for a game. We take the reviews for the most recent 3 years to align with the CoC collection timeline. 

Figure~\ref{fig:tox_forest} shows our results. We categorize games by their community sizes (total number of reviews). 
Our result shows that the availability of CoC is positively associated with player-perceived toxicity. 
Across all small to medium communities (< 10K reviews), 
games with CoCs exhibit significantly higher 
toxicity prevalence than games without CoCs 
(rank-biserial r = 0.20–0.35, p < 0.001). 
The association weakens and becomes non-significant 
for the largest communities (> 10K reviews), which is possibly due to the tension between an increased need of compliance versus the challenges in enforcing security and safety policies and shaping community culture~\cite{grace-chiplay22}.
This suggests that CoC adoption is responsive to community problems.

We additionally discover that CoC availability is significantly associated with a game's anti-cheat certification on Steam, an indicator of strong security enforcement. Among games with certified anti-cheat deployed, 23.5\% publish a CoC, compared to only 3.1\% among games without anti-cheat ($\chi^2$=268.4, $p<0.001$, $dof = 1$).
This suggests that  games without anti-cheat are also less likely to provide CoC.

\begin{figure}[t]
    \centering
    \includegraphics[width=\columnwidth]{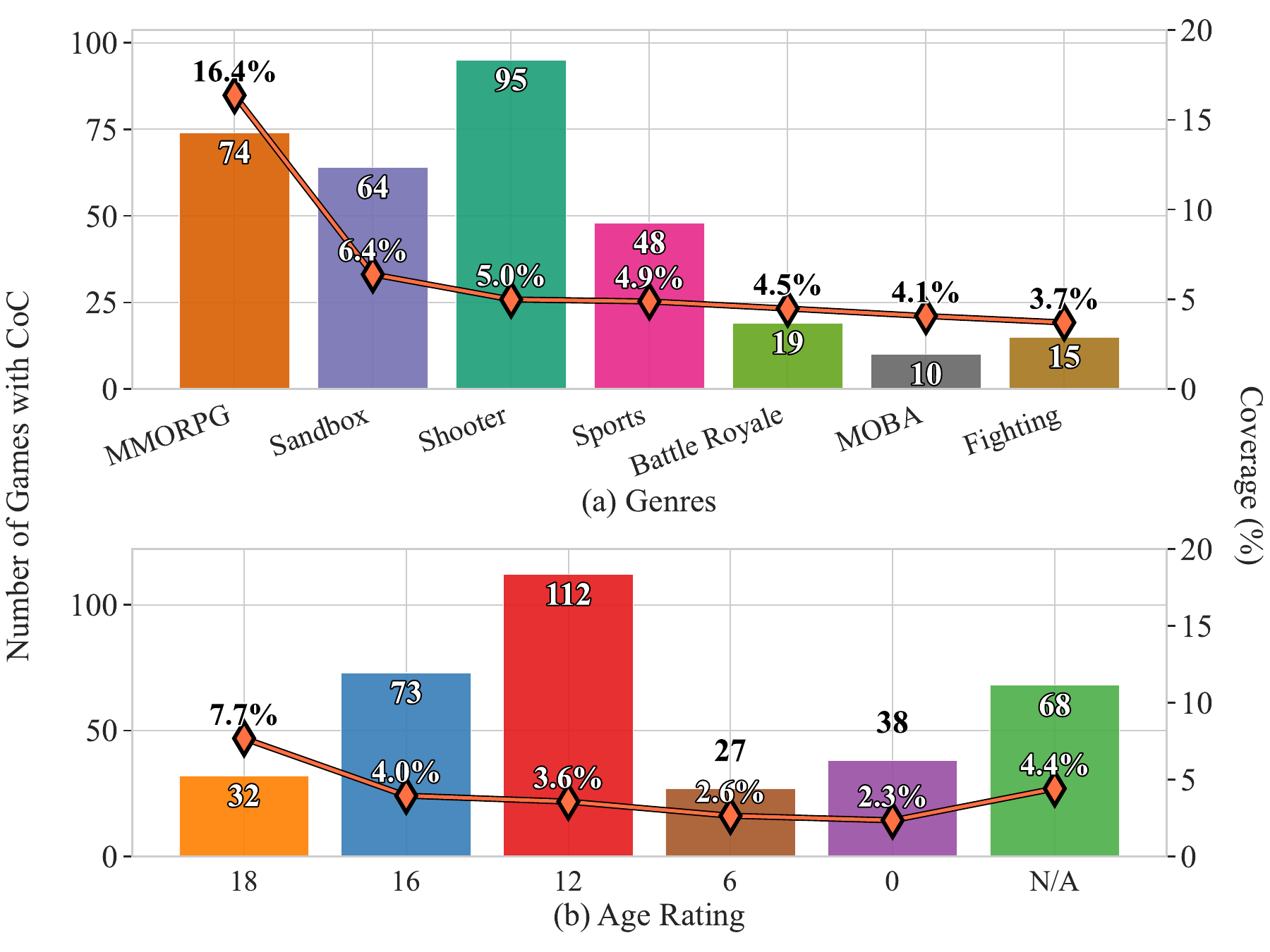}
    \caption{CoC availability frequency and ratio (\% of CoC-available games) per (a) genre and (b) age rating.} 
    \label{fig:adoption_wide}
\end{figure}

\noindent
\textbf{Community-centric \textit{MMORPG} titles exhibit the highest availability rate (16.4\%), compared to games that feature more intense gameplay.}
Figure~\ref{fig:adoption_wide} (a) compares CoC availability across game genres we introduce in Section~\ref{sec_Method_Dataset}. 
Note that a game may be associated with multiple genres.

CoC availability rates in most genres are relatively low.
The \textit{Shooter} genre, which often involves fast-paced interactions eliciting conflicts and player harms~\cite{to_2004_actiongame},  includes the highest number of games with CoC available in our dataset; however, compared to the large number of  \textit{Shooter} games in the marketplace, the availability rate remains low (5.0\%).
\textit{MMORPGs} are far more likely to publish a CoC: 16.4\%, compared to other titles.
This aligns with \textit{MMORPGs}' persistent, large-scale communities that require ongoing governance as well as highlights n identity development and reputation systems~\cite{to_2004_pvp, to_2004_adventure}.
This overall difference is significant ($\chi^2$=99.3, $p<0.001$, $dof = 6$).

\noindent
\textbf{CoC availability rates rise with age thresholds, suggesting gaps for a large number of under 18, child-appropriate games.} 
Figure~\ref{fig:adoption_wide}(b) shows that CoC availability rates increase from  2.3\% (38/1,619) in \textit{0+} to 7.7\% (32/417) in \textit{18+}.
The overall difference is significant, verified by $\chi^2$ test ($\chi^2$=33.04, $p<0.001$, $dof = 5$).
This may correspond to the observation that \textit{18+} games generally present a higher degree of toxicity.
However, many child-appropriate games determined by content rating still lack a CoC, failing the regulatory recommendations, even though children can face real harms (e.g., violence, sex) in player interactions~\cite{faraz2022child, zhang2025dangerous}, and underage players' voices may be suppressed among the player percetions online.

\subsection{Governing Violation (RQ2)}
\label{sec_Results_Misconduct}

Our analysis for \textbf{RQ2} below reveals to what extent CoCs' approaches differ in governing various forms of misconducts, including security, privacy, and safety violations.

\subsubsection{How relevant are different forms of misconduct to varying types of games}
\label{sec_Results_Misconduct_Coverage}
\begin{figure}[t]
    \centering
    \includegraphics[width=\columnwidth]{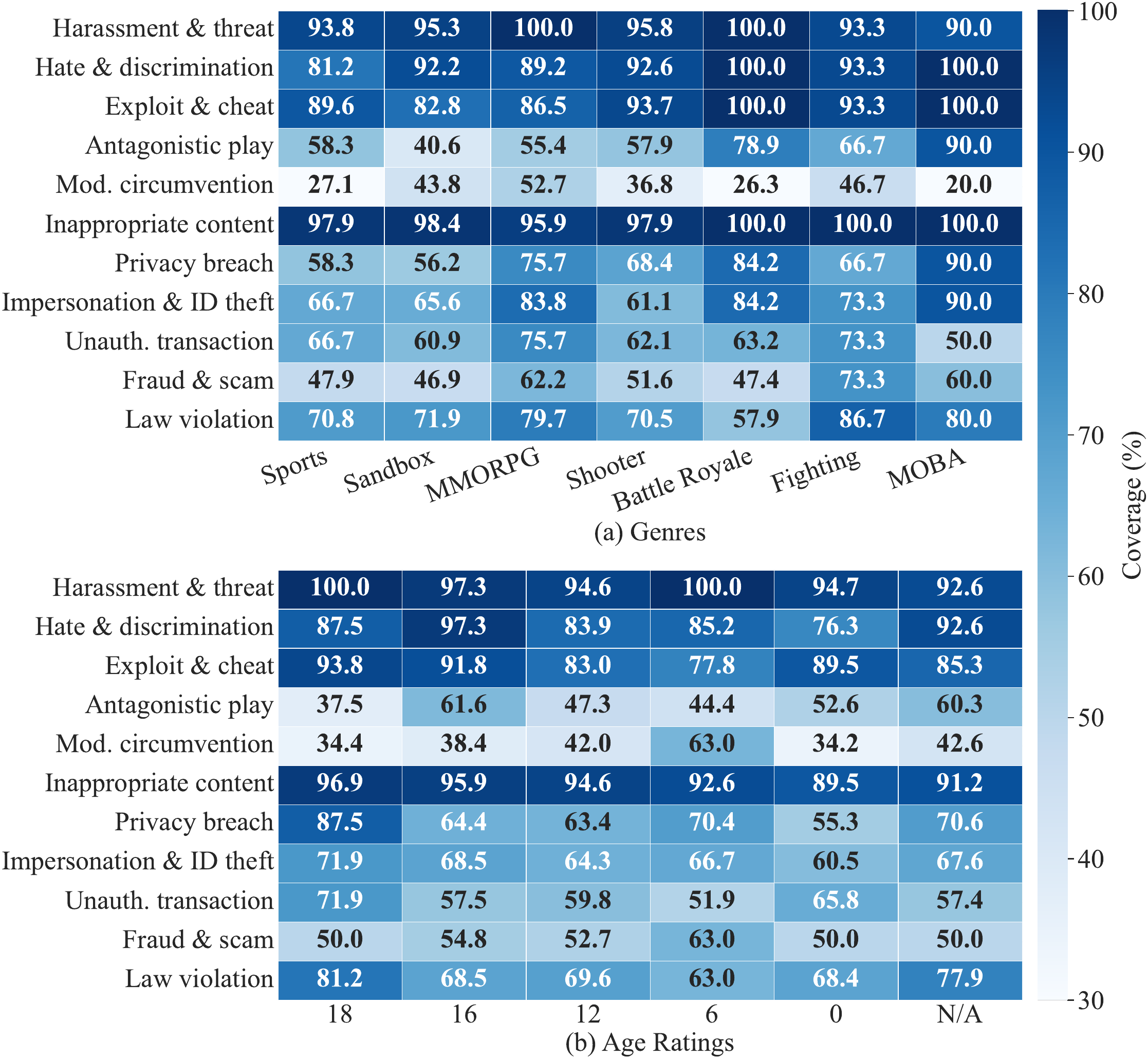}
    \caption{Coverage of misconduct labels across game attributes. Each panel shows per-game coverage (\%) by (a) genres and (b) age ratings, with darker cells indicating higher prevalence of the corresponding misconduct.} 
    \label{fig:Misconduct_wide}
\end{figure}

Figure~\ref{fig:Misconduct_wide} shows how the coverage rates of the 11 forms of online safety misconduct, as previously identified in Section~\ref{sec_framework_safetyanalisis}, in CoCs vary by age ratings and game genres. Our analysis shows key insights below.

\noindent
\textbf{Four consistently recognized forms of misconduct (covered by >80\% titles) align with the most salient player harms, while genre-specific differences emerge in less common ones.}
As shown in Figure~\ref{fig:Misconduct_wide} (a) and (b), above 80\% of games in different genres highlight four consistent misconducts: \textit{Harassment and threat} (90.0\%--100.0\%), \textit{Hate and discrimination}(81.2\%--100.0\%), \textit{Inappropriate content creation and sharing} (95.9\%--100.0\%), and \textit{Exploiting and cheating}(82.8\%--100.0\%), mirroring the top high-level risks reported by players~\cite{seng2019analyzing, zhang2025dangerous}.

In contrast, genre differences mainly appear in lower-coverage violations. 
\textit{MMORPG}, \textit{Fighting} and \textit{MOBA} games show higher coverage rates of a broader set of misconducts, 
e.g., \textit{Privacy breach} (75.7\% for \textit{MMORPG} and 84.2\% for \textit{Battle Royale}), \textit{Impersonation} (90.0\% in \textit{MOBA} and 84.2\% in \textit{Battle Royale}), and \textit{Law violation} (86.7\% in \textit{Fighting}).
The broader coverage of \textit{MMORPG} is consistent with their focus on role-play identities~\cite{to_2004_roleplaying}, as well as challenges driven by a more complex game ecosystem and economics (e.g., trading and high account value)~\cite{to_2004_mmo, to_2004_pvp} that motivate the focus on \textit{Impersonation and identity theft} and \textit{Unauthorized transaction and commercialization}.
The broader coverage of violations for \textit{Fighting}, \textit{MOBA}, and \textit{Battle Royale} are associated with their focus on ranking and frictions in competitive gameplay; however, this higher coverage may just represent the safety posture of a relatively smaller number of games ($<$20 each) that have CoC available in these three genres, as compared to the 95 \textit{Shooter} games which share similar frictions due to competition.

We use permutation-based $\chi^2$ tests with false discovery rate (FDR) correction to evaluate associations between misconduct coverage frequencies with game genre, which does not show strong association overall.
This echoes prior user research that game genres, as interaction contexts \textit{by design}, are not the only drivers of safety and security violations, and community development plays a critical role, in addition to the correlations that exists across games that share multiple genres~\cite{zhang_moradzadeh_woan_kou_2024}.

\noindent
\textbf{\textit{18+} games recognize a broader range of misconducts.}
\textit{18+} games show higher coverage rates for a broader set of misconducts, which tops 7 of the 11 forms of misconducts, highlighting the overall tension in gameplay.
Although we do not observe strong signals of association aggregated at the game level, segment-level co-occurrence shows significant associations between age rating and misconduct types (age rating: $\chi^2=83.20$, $p = 2.00 \times 10^{-3}$, $dof=50$).

\noindent
\textbf{Coverage of high-level violations does not reflect player-perceived toxicity of individual games.}
Our further examination of the association between player-perceived toxicity and coverage of high-level violation does not show significance using Spearman correlation tests ($r=0.046, p=0.390$).
This contrasts our earlier finding that more-toxic games 
are more likely to publish a CoC in the first place: while the decision to publish a CoC tracks community toxicity, there is not a single form of violation that stands out.

\subsubsection{How specific are governance strategies against different types of misconduct} 
\label{sec_Results_Moderation_Misconduct}
\begin{figure}[t]
    \centering
    \includegraphics[width=\columnwidth]{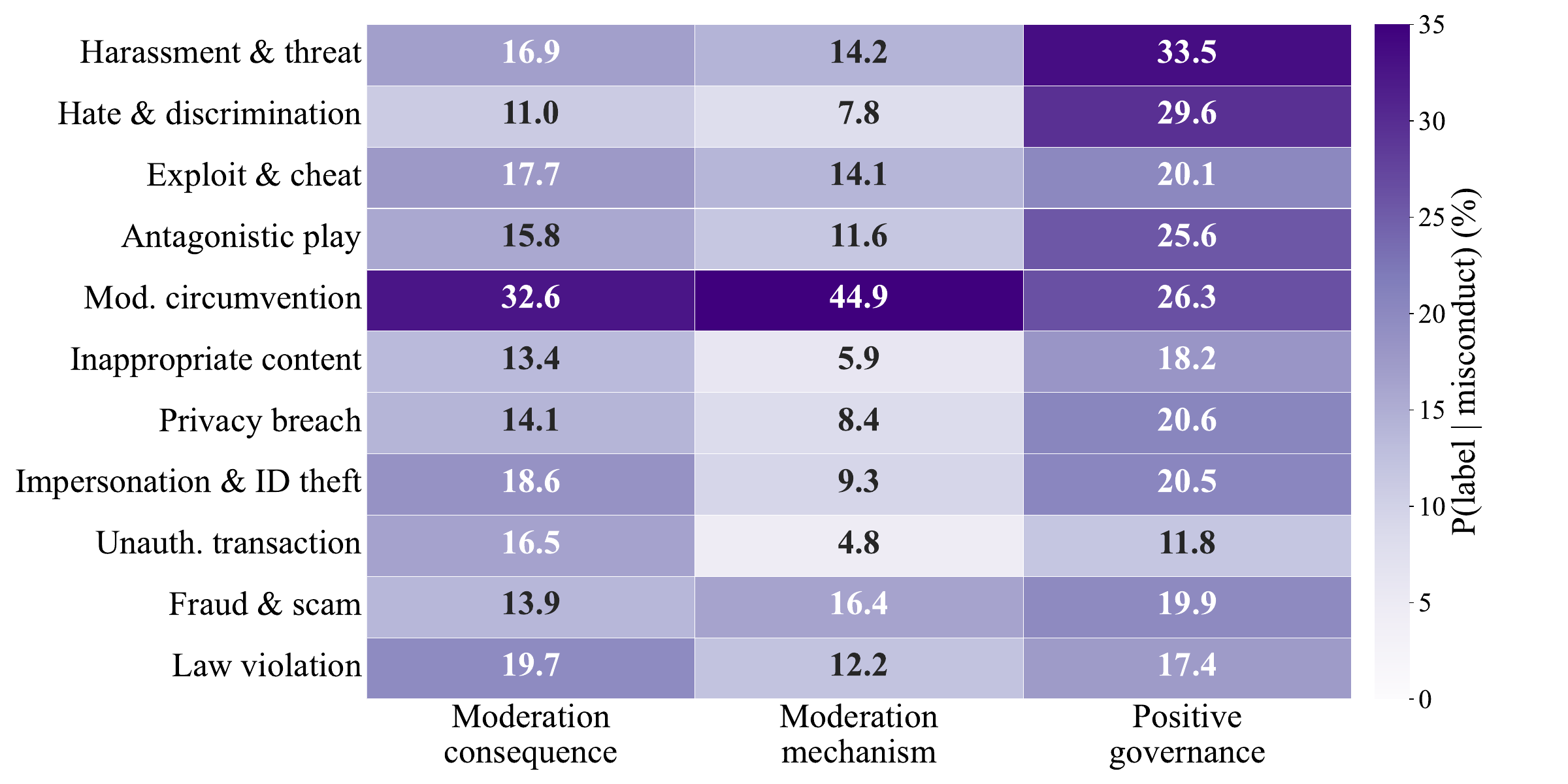}
    \caption{Segment co-occurrence rates between misconduct types (rows) and governance strategies (columns), with darker cells indicate misconducts that are more often co-locate with concrete moderation framing or positive-governance language.} 
    \label{fig:misconduct-moderation-row-norm}
\end{figure}

Figure~\ref{fig:Moderation_wide} in Appendix shows that coverage rates of governance strategies, specifically consequences of detected violations, moderation mechanisms, and positive communication, in a CoC, are overall high (above 81.2\%) across genres and age ratings, 
while genres show slightly larger variations (81.2\%–100.0\%).
We further measure the associations between governance strategies and specific misconduct types using their segment-level co-occurrence (Figure~\ref{fig:misconduct-moderation-row-norm}).

\noindent
\textbf{CoCs overall highlight punishments over misconduct-specific moderation mechanisms, reflecting gaps in enforcement.}
Overall, misconduct types and moderation strategies are strongly associated 
($\chi^2=67.27$, $p<0.001$, $dof = 10$).
CoCs more frequently mention the consequences of a violation rather than the mechanisms of moderation. 
\textit{Unauthorized transaction and commercialization} shows the largest gap (16.5\% vs. 4.8\%), likely because such violations often involve off-platform side channels (e.g., real-money trading) where enforcement is challenging. For \textit{Inappropriate content creation and sharing}, specific \textit{moderation mechanisms} are also rarely associated with a misconduct provision (5.9\%), as these cases could be often handled through general reporting.
In contrast, \textit{Circumventing and abusing moderation mechanism} are more often discussed with \textit{Moderation mechanisms}, which naturally connects to the nature of the this abuse.

\noindent
\textbf{Positive governance is inconsistently applied against different violations.}
Compared to punitive moderation analyzed above, positive communication in CoC serves as a proactive nudge to guide player behavior, which is recommended by policy-makers and practitioners~\cite{EU_DSA, FPA_Framework}. 
These communications provide alternatives to mitigate violations proactively, such as support team members, and highlight community values to rationalize safety and security policies, e.g., inclusivity~\cite{anykey_inclusion}.
However, this strategy is inconsistently applied on violations that involve harms that arise from interpersonal interactions and playing (\textit{Harassment and threat}, \textit{Hate and discrimination}, and \textit{Abuse of play}); while CoC's sentiment on violations against the game service itself is more punitive.

\subsection{Context Specificity (RQ3)}
\label{sec_results_Moderation}

As described in Section~\ref{sec_methodology}, we leverage \workname to measure to what extent a CoC contains entities that are specific to the gaming and safety contexts of individual games, as compared to the entire corpus. 
In \textbf{RQ3}, we first use this as a metric to assess (1) what types of game are more likely to offer specific CoCs, (2) how context-specific are CoCs when defining different violations, and (3) how different gaming and safety contexts make CoCs distinct.

\begin{figure}[t]
    \centering
    \includegraphics[width=\columnwidth]{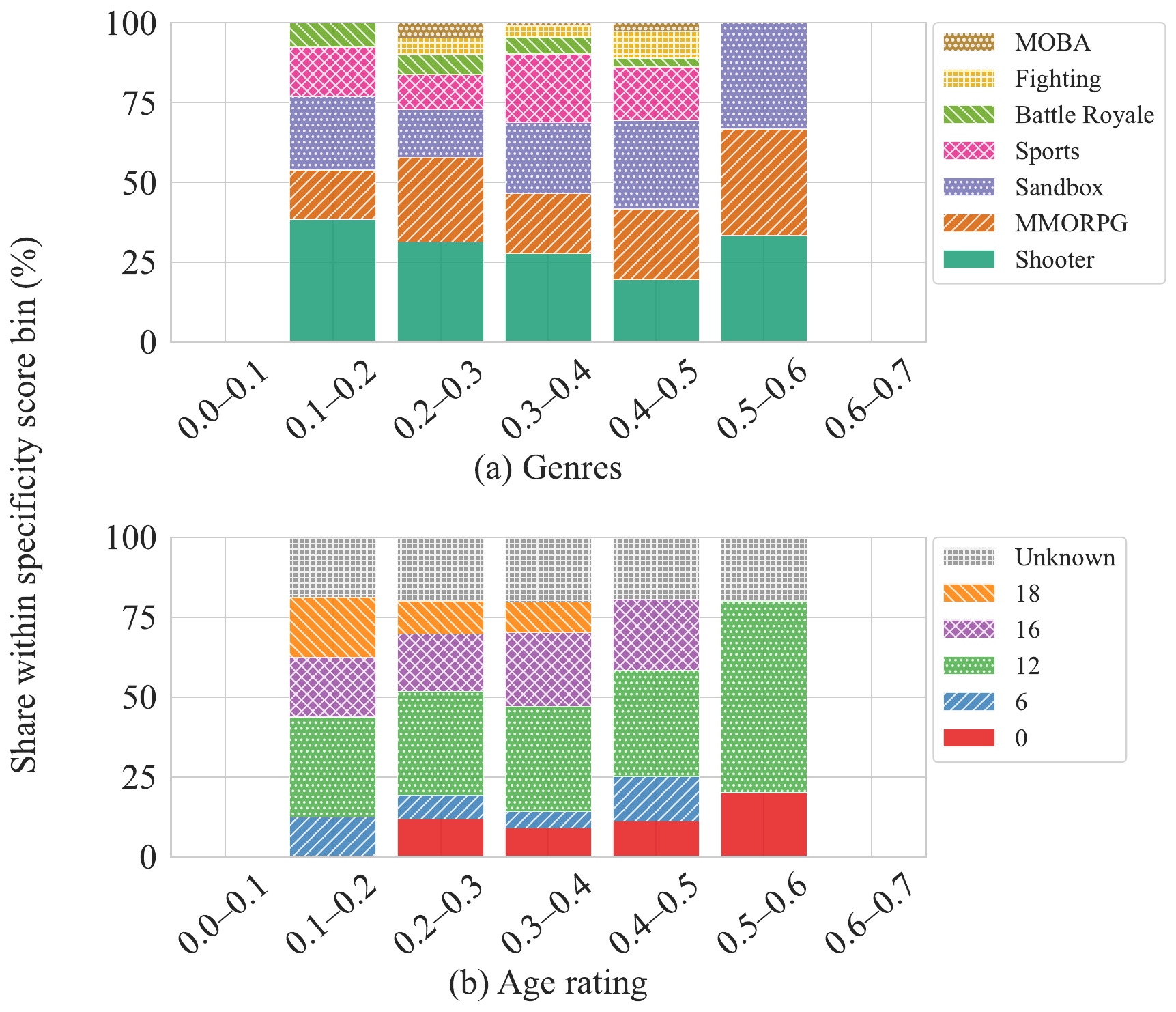}
    \caption{Normalized distributions of distinctiveness scores by genre and age rating.}
    \label{fig:stackdist}
\end{figure}

\subsubsection{What types of game are more likely to offer specific CoCs}
\label{sec_Results_Specificity_Type}

Figure~\ref{fig:stackdist} shows distribution of specificity score for games in different genres and age ratings. 
Following the modified Hausdorff distance defined in Section~\ref{sec_Method_Key_Entity}, we measure specificity as the mean nearest-neighbor distance between a game's local topic centroids and the centroids of the global topic model:
$
\ell(\mathcal{T}_g, \mathcal{T}_G) 
= 1 - \frac{1}{K_g}
\sum_{\mathbf{c}^g \in \mathcal{T}_g}
\max_{\mathbf{c}^G \in \mathcal{T}_G}
\mathrm{sim}(\mathbf{c}^g, \mathbf{c}^G)
$. Higher values indicate more context-specific governance language relative to corpus-level governance patterns. We reveal the following findings.

\noindent
\textbf{The CoC content of \textit{18+} games is least 
context-specific.} Among the six age rating categories, \textit{18+} 
games show a monotonically declining share across distinctiveness 
bins (18.75\% at 0.1–0.2, 10.34\% at 0.2–0.3, 9.70\% at 0.3–0.4, 
and 0\% at 0.4–0.5 and beyond). \textit{16+} and \textit{6+} games 
similarly disappear from the highest distinctiveness bin (0.5–0.6), 
while \textit{12+} games maintain a substantial within-bin share 
across all bins (31.25–33.33\% in the 0.1–0.5 range) and account 
for 60\% of the highest distinctiveness bin alone. \textit{12+} 
games, also the most numerous category in our sample, therefore 
span the full distinctiveness range, whereas \textit{18+} CoCs 
cluster in standardized, low-distinctiveness language.

\noindent
\textbf{\textit{MMORPG} and \textit{Sandbox} games offer more 
distinct contexts in CoCs.} We observe that the two game genres, \textit{MMORPG} and \textit{Sandbox}, that features narrative, open-world, and creativity in the gaming mechanics~\cite{to_2004_mmo, to_2004_sandbox} show better diversity when contextualizing their safety policies, which increase their combined within-bin share as distinctiveness rises (combined around 38\% at 0.1–0.2, around 50\% at 0.4–0.5, and about 67\% at 0.5–0.6), while \textit{Battle Royale}, \textit{Fighting}, 
\textit{MOBA}, and \textit{Sports} drop to 0\% there.

\begin{figure}[t]
    \centering
    \includegraphics[width=\columnwidth]{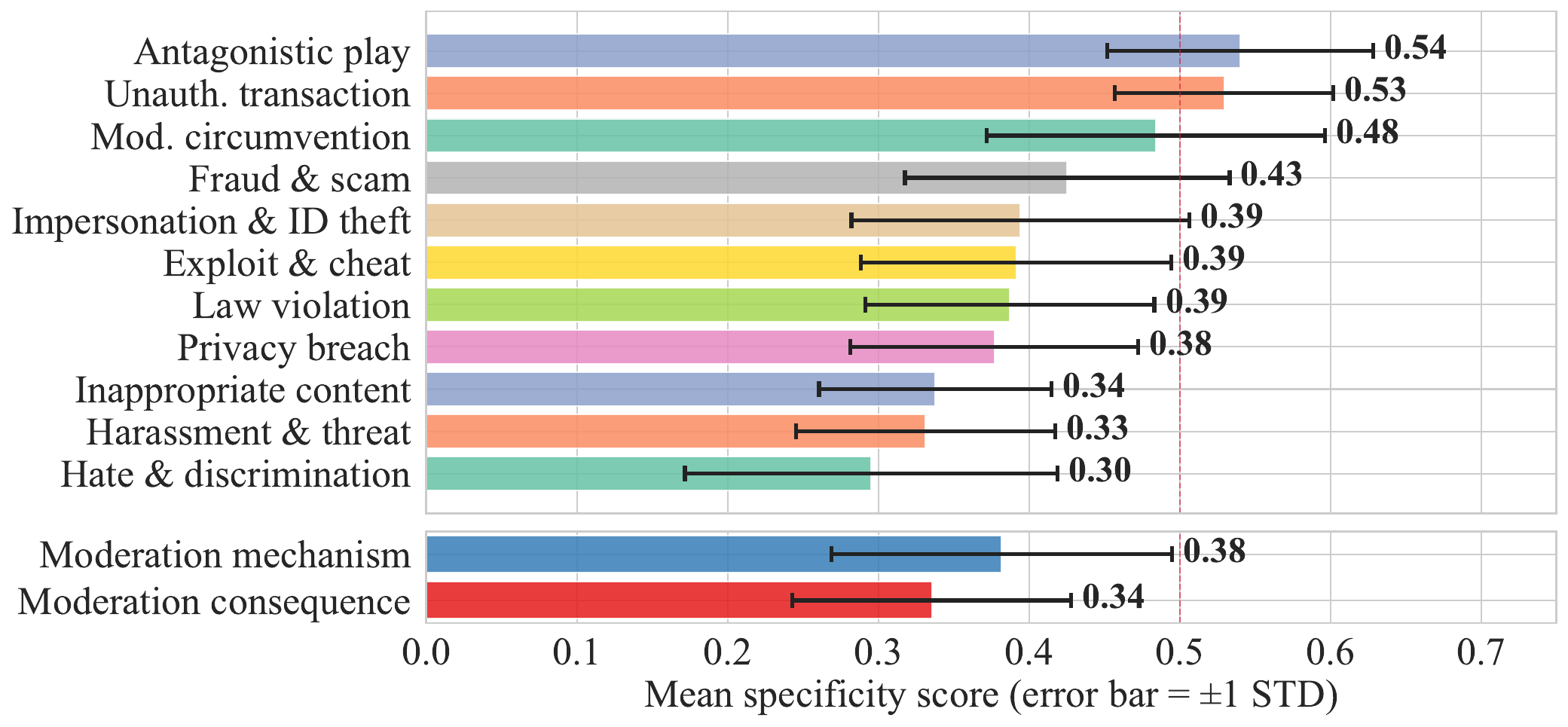}
    \caption{Per-label distribution of  distinctiveness scores.} 
    \label{fig:label-distinct}
\end{figure}

\subsubsection{How context-specific are CoCs when defining different violations} 
\label{sec_Results_Moderation_context}

Figure~\ref{fig:label-distinct} shows the distribution of per-game distinctiveness scores filtered and categorized by segments on 11 misconduct and two moderation topics. The result highlights the following.

\noindent
\textbf{Game-mechanic-related violations are more contextualized, 
while interpersonal harms are less distinct.}
First, \textit{Hate and discrimination} and 
\textit{Harassment and threat} show the lowest mean 
distinctiveness, suggesting that these categories largely share vocabulary, closely followed by \textit{Inappropriate content 
creation and sharing}. In contrast, 
\textit{Abuse of play}, 
\textit{Unauthorized transactions}, 
and \textit{Circumventing and abusing moderation} show 
the highest means, reflecting that these violations are discussed 
under complicated contexts. In addition, compared to definitions of misconduct, the descriptions for moderation-related content are relatively uniform.

\subsubsection{How different gaming and safety contexts make CoCs distinct}
\label{sec_Results_Moderation_Specify}
Through contextual entity extraction and analysis (Section~\ref{sec_Method_Key_Entity}), we examine how CoCs specify online safety violations and moderation processes in ways that are shaped by gameplay mechanics and community structures. 
We then discuss the major themes surrounding these contextual entities. 
Building on the topic modeling and specificity measurements above, we highlight the key categories of contextual entities , identified from topic modeling and indicated with underlining, that anchor our content analysis.

\noindent
\textbf{Who are CoCs protecting: CoCs highlight community roles and relationships, but considerations of protected demographics remain superficial.}
CoCs frequently specify the potential targets of interpersonal harms such as harassment. Beyond referring to players or users collectively, many CoCs enumerate protected \underline{demographics}, such as gender, religion, race, and sexual orientation, reflecting the inclusive values upheld by gaming communities. However, such specification can remain relatively surface-level: CoCs often list protected characteristics without considering the most active or vulnerable audiences within a given community, including children. This helps explain the observation in Figure~\ref{fig:label-distinct} that interpersonal harms are among the least distinct categories.

Using a lexicon-based approach (the complete search terms are available in Appendix~\ref{keyword_list_lexicon}), we find that only 24\% of the games studied mention child-related content in their CoCs. Moreover, the association between age rating and mentions of children is not significant ($\chi^2=4.49$, $p=0.34$, $dof = 4$). Notably, many references to content that is \textit{``inappropriate for children''} (EverQuest, \textit{MMO}) frame children primarily in relation to content exposure risks, rather than as underage players who may be directly involved in social interactions.

By contrast, some CoCs specify protected identities through community structures and social ties with greater contextual detail. First, CoCs often protect the \uline{communities and teams} with which players are affiliated, treating groups such as \textit{clans} as collective units that can be harmed. This is particularly salient in \textit{MMORPGs}, where player groups are embedded in gameplay progression and social organisation. For example, Idle Clans (\textit{MMORPG}) prohibits \textit{``stealing items from a clan vault or not respecting the accepted terms in group boss fight runs when it comes to loot distribution.''}

Second, more specific CoCs extend protection to \uline{practitioners in the wider ecosystem}. For example, Splitgate 2, a competitive \textit{shooter}, bans players from impersonating \textit{celebrities} and \textit{streamers}, recognising that impersonation and harassment can extend beyond in-match interactions to the wider attention economy surrounding competitive games.

Third, \uline{game masters and moderators} are explicitly named as targets of harmful behavior by some games. For instance, Myth of Empires, a multiplayer war sandbox, refers to \textit{``Game Masters or Myth of Empires employees''}. Because these actors mediate conflicts and oversee fair play, they occupy a visible and potentially vulnerable position of authority within the community.
Additionally, CoCs protect \underline{companies and staff}, banning broader abuse toward developers and partners.

\noindent
\textbf{What content is inappropriate: extending content governance from age suitability to unauthorized creative, commercial, and fraudulent content in broader ecosystem.}
More than describing inappropriateness as restriction of age ratings, we find that ``inappropriateness'' can be governed as a boundary regulating unauthorized creative, commercial, and fraudulent activity within game communities in more specific ways by games.
In a more game specific way than enumerating types of \underline{harmful information} by categories already restricted by age ratings, CoCs may discuss \textit{disinformation} and \textit{exploits} that can harm gameplay integrity.
Furthermore, some CoCs flag specific ways of \underline{privacy and copyright} violation by sharing \textit{account information} and leaking \textit{copyrighted content} or spoilers. Off The Grid, an online \textit{Battle Royal} game, states \textit{``Do not engage in account sharing''}. Smash Legends (an anime-style \textit{fighting} game) bans players from using \textit{``A nickname that may infringe on the trademark or copyright of a third party''}.
Additionally, CoCs regulate \underline{advertisement and spam} by separating community promotion from commercial activities, e.g., promoting guild and fan groups for \textit{``any type of financial or commercial gain''} is a violation (Wolfenstein: Youngblood, a \textit{18+ Shooter} game). 
Finally, CoCs regulate the \underline{information channels} where harmful content appears: not only the regular text and images, but also game room names, logos, and avatars, demonstrating the interaction contexts in game.

\noindent
\textbf{What can be exploited and how: concerns of abusing accounts, gameplay and technical systems center around fair play.}
We discover that the vulnerabilities and exploits concerned by CoCs centered around fair play showing large degree of specificity.
First, the abuse of \underline{accounts and game characters} is among the most consistent concern, as practices such as \textit{``manipulating additional accounts''}) undermine fair progression and role consistency (Smash Legends, a anime-style \textit{Battle Royale}).

Second, CoCs regulate misconducts associated with \underline{gameplay design and playstyles}, addressing behaviors such as \textit{teaming},  \textit{ranking} manipulation, or disruptive practices such as \textit{trolling}.
A recurrent focus is the abuse of game economies, particularly through real-world \textit{currency} trading as well as \textit{``abnormally moving gold (Rupiahs), in-game goods, etc. between specific accounts...''} (Trimurti Online, an \textit{MMORPG})

Third, CoCs ban \underline{technical cheating}, where some exploits such as ``using bots, hacks, macros, sandboxes or any other third-party software'' are commonly applicable across various types of games (NosTale, an anime \textit{MMORPG}). 
Other games discuss exploits that are more relevant to its gameplay and system. 
For World of Warships, a naval simulation game, this includes abusing core systems, e.g., \textit{``intentionally exploiting the physics system''}, where physics underpins realism and fair combat. In Call of Duty: Infinite Warfare, a first person shooter game, CoCs instead enumerate specific tools, such as \textit{``aimbots, wallhacks, trainers, stats hacks, texture hacks, leaderboard hacks, injectors, or any other software used to deliberately modify game data on disk or in memory.''}

\noindent
\textbf{Who are responsible for enforcing moderation: layers of moderation associated with community and organizational hierarchies.}
We find four actor groups collectively forming an escalation ladder: users flag issues, moderators adjudicate, companies govern outcomes, and external authorities handle the most severe cases.
\ul{Players and users} are most mentioned who serve as frontline actors, who are expected to report, but are also explicitly discouraged from \textit{"naming-and-shaming other users"} (\textit{Wolfenstein Youngblood, a first-person shooter}). 
Some CoCs also extend accountability to guardians via \textit{``Parental Consent''} (\textit{Synth Riders}, a VR rhythm game).
Meanwhile, \ul{moderators} serve as the adjudication layer, who collect reports, mediate conflicts, and apply sanctions. Yet, their authority is often scoped by platform contexts: e.g., VRChat defines a boundary between \textit{``Group moderation''} and the central \textit{``VRChat Trust and Safety.''}.
Games occasionally mention  their \ul{game companies and staff} overseeing the process and may escalate cases to \ul{external authorities}. When harms cross legal or offline boundaries, companies may contact \textit{``local law enforcement''} (EVE Online, a spaceship \textit{MMO}).

\noindent
\textbf{What is the process of moderation: more specific communication of player-facing procedures than back-end mechanisms.}
The discussions of moderation mechanisms are often less clearly defined.
Three mechanisms recur: \ul{player reporting}, \ul{in-game detection and analytics}, and \ul{complaints and appeals}.
Some CoCs may provide explicit instructions for player reporting, such as to  \textit{``insert the avatar link in the 'Offending URL' section"}  to submit evidence (Battlefield Hardline, an \textit{Action PvP})
For \textit{in-game detection and analytics}, CoCs sometimes cite log- and stats-based signals, which are useful for quantifiable harms (e.g., team damage), e.g.  \textit{``actively scans the game logs searching for players who use automated bots''} (Fishing Planet, a fishing simulator).
Many CoCs acknowledge dissatisfaction and offer channels for appeals, e.g., \textit{``appeal a moderation decision applied to your server''} (Unturned, a survivor game).

\noindent
\textbf{What is the consequence of violation: punishments focus on account and content restrictions with varying levels of specificity.}
We identify 3 major types of consequences for a CoC violation: warnings, content or communication limits, and account-level suspension or bans.
\ul{Suspensions and bans} dominate, and are often account-wide, though some forms of suspension target features closely tied to problematic behaviors, (e.g., \textit{``features that are most closely associated with the problematic behavior''} (Ara: History Untold, a turn-based strategy) and such as \textit{`` in-game player data such as inventory, owned ships''} (Starbase, a space MMO), with durations sometimes specified.
Beyond restrictions, CoCs may enforce \ul{content or communication restrictions}, such as removing or editing posts, or muting players who violate rules.
In some cases, CoCs adopt a staged approach, issuing \ul{warnings} for \textit{``corrective actions''} before escalating to harsher penalties (WWE 2K25, a wrestling game).

\section{Discussion}
Our study sheds light on core questions concerning digital harm and online safety governance in the video game industry.
We identify the current gaps to addressing complex socio-technical harms under the CoC framework, meeting the expectations of the players and policymakers, and future opportunities to improve safety governance.

\noindent
\textbf{Should more games adopt CoC?}
Section~\ref{sec_Results_Adoption} shows that standalone CoCs remains rare (4\%) compared to other commonly implemented web documents such as Terms of Service~\cite{sundareswara2021large} or  privacy policies~\cite{wilson2016creation, harkous2018polisis, amos2021privacy}, although the availability of CoC is assocaited with reduced toxicity levels. 
CoCs are far less ubiquitous, especially among games outside major studios, leaving many platforms in the so-called ``long-tail'' without clear behavioral guidelines for community behavior and recourse for misconduct.
The lack of availability in certain games might be justified by the relatively limited user interaction and weak sense of community belonging. 
For example, Section~\ref{sec_Results_Adoption} discovered that video games with a stronger community and narrative focuses (e.g., \textit{MMORPG}) tend to have higher CoC availability and mention more concrete safety practices (Section~\ref{sec_Results_Misconduct_Coverage}). 
This raises the question of whether regulators should mandate CoCs for games that involve substantial user interaction.
Such mandates could increase CoC adoption, including among smaller platforms
However, they may also intensify the tension between game development and social responsibility by imposing additional compliance and moderation costs, particularly on small or independent studios with limited resources.

An alternative view argues CoCs may emerge from bottom-up processes. Encouraging developers to see CoCs as improving player experience and retention may be more effective, especially given their flexibility in defining rules and enforcement. 
Learning from the proactive figure of  \textit{MMORPG} titles we observed, community and identity-building might be a good strategy to strengthen online safety from bottom-up.

\noindent
\textbf{For the vendor or the players?}
CoCs serve both vendors and players, but the balance between these interests is often unclear.
Some provisions directly protect the vendor.
Discouraging exploits through a CoC can be more cost-effective than investing in technical measures, which divert developer efforts away from game development. 
Other provisions more directly protect players, such as preventing harassment, hate and discrimination, and fraud.
Such safety and security harms are technically challenging to mitigate, especially when they occur out-of-channel (Section~\ref{sec_Results_Moderation_Misconduct}), so users are asked to follow safety norms as a soft measure.
Game vendors therefore have incentives to reduce harmful incidents that degrade user experience by shaping player behavior, as shown in Section~\ref{sec_Results_Moderation_Specify}.
However, from players' perspectives, the current design of CoCs provide limited transparency into the the safety and security status quo within a game, other than a few platform enforcement such as certified anti-cheating.
This tension is also visible in violations such as harassment and hate, which are among vendors' central concerns.
Most CoC address them using standardized language, such as enumerating protected characteristics. 
Such language may be important for inclusivity and legal completeness, but it often does not reflect the specific risks, audiences, or social dynamics of a particular game community.
For example, as shown earlier, CoC content does not noticeably adapt to minors when they are part of the user base. 
Nevertheless, gameplay-related rules, which exhibit a higher level of diversity in acknowledging context-specific harms and exploits (Section~\ref{sec_Results_Moderation_context}), although the engagement and interactions between gameplay-specific exploits to interpersonal harms can be improved for players. 

A further limitation is that CoCs rarely make clear how strongly vendors are committed to enforcement and how effective the enforcement is. 
Although many CoCs provide reporting channels, they often offer little information about specific enforcement thresholds, response procedures, or sanction consistency. 
This opacity matters because vendors may face conflicting incentives. 
For instance, a game or community may hesitate to enforce anti-discrimination policies that protect marginalized players if doing so risks backlash from other segments of the player base.
Ideally, such cases would be rare, but the experiences of women in gaming suggest that this concern should not be ignored~\cite{bryterglobal2023women}.
Even if CoCs were mandated, vendors that are reluctant to enforce player-protective rules may still implement them only superficially.
One avenue is for regulators to extend the video game governance framework from content safety to behavioral safety, which is central to players' lived experience of online games.
Regulatory frameworks have historically focused more on content classification, age ratings, and consumer protection than on the governance of player-to-player conduct.
Regulators therefore could draw on transparency-oriented frameworks, such as EU DSA~\cite{EU_DSA}, to require greater visibility into reporting systems, enforcement practices, and safety outcomes.

\noindent 
\textbf{Improving CoC integration.}
Developers seeking to create a CoC need resources to produce policies that address the right safety issues in a way that users from different backgrounds and age groups can understand, which is also demanded by online safety regulators~\cite{EU_DSA, UKOnlineSafetyAct}.
This may not be straightforward for online safety violations and measures that are technical or blurry in nature (Section~\ref{sec_Results_Moderation_Specify}), and full transparency might increase the risks moderation get circumvented.
While sharing template CoCs may seem appealing, as in privacy compliance~\cite{amos2021privacy}, our results suggest this approach is sub-optimal: CoCs may vary substantially by game objectives, play style, and audience, with no clear standard by features such as genre we know at design time.

A promising direction is to develop toolkits that help developers identify the specific risks and needs relevant to their particular games, audiences, and broader platform ecosystems.
Ideally this process would involve players and community leaders through co-design methods, or incorporate consent-based analyses of player feedback, building on \workname.
Such developer-oriented tools could take inspiration from privacy compliance systems, such as automated privacy policy generators~\cite{pan2024trap}.
Rather than providing one-size-fits-all templates, these toolkits could offer structured processes, structured processes, best practice examples, and checklists that help developers craft CoCs.
In addition, LLM-based persona simulation may help game vendors evaluate whether proposed CoCs and policies are understandable, inclusive, and responsive to the needs of different player groups~\cite{li2025makes}.

\noindent
\textbf{Future Work.}
In addition to the opportunities described above, we further outline a few broader venues for future studies.
First, we discovered the association between player-perceived toxicity and CoC and the limitations for CoC to address needs of specific player groups. We argue that measuring the effectiveness of CoCs with particular player groups, as well as the influencing factors and real-time status of CoC enforcement, is challenging, due to sample availability and restrictions in studying proprietary software, yet valuable. Future work could invite players and moderators from different backgrounds such as parents~\cite{liu2025co} in studies and pilots of using CoCs and moderation tools.  

Second, while we focus on Steam, \workname can be easily adapted to study other platforms, for example, Google Play~\cite{googleAndroidApps}, to understanding safety issues in mobile games, or more bottom-up player communities. 
This will open up interesting questions, as the governance approaches on these platforms differ.
Steam is known to take a more libertarian approach in profiting, which potentially influences its attitude in safety governance~\cite{ftValveConquered}. 
For example, age rating is not mandated on Steam for many countries unlike Google.
Meanwhile, it is interesting to conduct a longitudinal analysis and understand how changes of CoCs are driven by the development of safety regulations such as EU's Digital Service Act and UK's Online Safety Act~\cite{law-jml24} or drifts in player community cultures~\cite{saldanha-gc23}.
Additionally, future work can cross-compare CoCs with other resources, such as player forum discussions, to provide a more holistic view of the safety issues embedded in the community~\cite{winter-chi25, vetrivel-sec23, li-sp23}. 

Last, we believe \workname provides a foundation for future NLP research on safety policies, inspired by the series of academic developments in automated privacy policy analysis.
For instance, a safety ontology can be built to accurately break down a CoC rule for more fine-grained analysis such as consistency between CoCs and their actual safety supports in game, taking the linguistic diversity and specific in-game vocabularies into account~\cite{wilson2016creation, andow2020actions, cui2023poligraph}.

\section{Conclusion}
Video games present unique challenges for online safety governance, as traditional measures such as pre-release content ratings only partially address risks arising from player interactions. To bridge this gap, game communities publish CoCs as a novel form of security policy to set behavioral expectations and provide safety resources, yet CoCs remain underexplored as a governance mechanism.
We introduce \workname, an NLP-powered framework for large-scale CoC collection and analysis. Our study shows several gaps in CoC adoption. CoC availability is skewed toward high-profile titles. While some safety concerns are common across games, they are inconsistently governed. 
Community-focused games address more specific violations, with substantial variation in the specificity when defining violations and providing moderation resources. 
CoCs lack specificity that align with the games' targeted audience, including children.
Together, our findings highlight new opportunities to evaluate and improve CoC-based governance for more effective player protection.

\clearpage
\newpage

\section*{Ethical Considerations}
We carefully considered the ethics implications below. 

\noindent
\textbf{Considerations for stakeholders.} Our research is most relevant to video game players and designers, game distribution platforms, regulatory bodies and researchers for online safety. 
Our research aims to \textit{benefit} all these stakeholders by providing recommendations for video game designers and distribution platforms to improve online safety governance and communication. For players and regulators, our research will inform them of the potential challenges and risks in safety governance. 
The most concrete benefit would come from improved governance on the games that they play.
Our research aligns with the online safety regulations proposed by multiple countries. 
For researchers, our open-source system will foster future work to improve online safety.
Our research analyzes and suggests improvements for online safety governance in a broad community, including players from different backgrounds and developers at different sizes, ensuring the \textit{justice} principle.
We consider \textit{respect for persons} broadly and respect the communities, games, and players' perspectives in our analysis, acknowledging their efforts made in online governance and player protection as well as understanding the realistic challenges they face.
Our online measurement complies with the regulation in our country that permits text mining for research, showing \textit{respect for law and public interest}.
Furthermore, our study that analyzes public player reviews is approved by our institution’s ethics review process.

\noindent
\textbf{Harms and mitigation.} 
The first potential harm concerns the use and analysis of player reviews on Steam. 
Though the data is public available and was accessed through Steam's official API, we acknowledge that players might not expect their data being analyzed for specific academic research purposes. 
To minimize the negative impact to individuals, our data collection and analysis did not include personal information, and we only report aggregate quantitative statistics to characterize the game reviews as a whole.
In respect of users' privacy, we choose not to release the player reviews collected.

Another potential harm from our study relates to the game developer's infrastructure and also their reputation.
Our country permits automated crawling and analysis of publicly available data for non-commercial research.
We mitigated infrastructure risk by rate-limiting our crawler with an at-least 10 second interval between page requests to mitigate the risk that it overloads the service.
Further, we directly quoted games to avoid mischaratcerizing them, which might have reputational consequences.

\noindent
\textbf{Decisions.} We judged that the risks were relatively properly mitigated, far out weighed by the benefits to improve online safety governance. To maximize our research benefits, we will share our research output with relevant stakeholders and make our research artifact available under appropriate license.

\section*{Open Science}
To promote open science and support future research, we will make our research artifact public, including the \workname measurement pipeline, CoC annotations with the labeling scheme, our trained model, and data access.



\cleardoublepage

\bibliographystyle{plain}
\bibliography{bibs/cdf2025,bibs/cdf2026jan,bibs/cdf2026may}

\cleardoublepage
\appendix
\appendix
\section{Appendix}
~\label{sec_appendix}

\subsection{Labeling Scheme for Reference Dataset}
~\label{sec_appendix-label}

\begin{squishitemize}
    \item \textsf{Misconduct}: The behaviors which are not accepted and do not align with the norms and rules of the game, its community and service provider.
    \begin{squishitemize}
        \item \textsf{Hate and discrimination}: Verbal or other type of abuses, including intimidation, ridicule, or insulting remarks, particularly based on another individual’s or group's actual or perceived identity, e.g., race, religion, color, gender, gender identity, etc.
        \item \textsf{Harassment and threat}: Unwanted behaviors and abuses that offend, harm, intimidate, coerce or oppress another individual or group.
        \item \textsf{Abuse of play and antagonistic play}: Direct and intentional disruption to the normative flow of gameplay or antithetical to the game's spirit, such as trolling, griefing, as well as pestering, bothering, annoying, griefing or otherwise inhibiting another player’s reasonable enjoyment of the game
        \item \textsf{Exploiting and cheating}: Manipulating or abusing the (game) system or using third party systems to gain an unfair advantage over others 
        \item \textsf{Inappropriate content creation and sharing}: The creation and sharing of information, content, or speech that are deemed uninvited and inappropriate, e.g., hateful content, spam, etc.
        \item \textsf{Privacy breach}: Invading another individual's digital or physical privacy, including unauthorized collection, use, sharing, or exposure of personal or sensitive information, such as name, address, account details, or any other identifiable data.
        \item \textsf{Impersonation and identity theft}: Any form of identity deception or misuse, including impersonating other players, moderator, notable figures, and service provider for purposes such as obtaining unauthorized access to another individual's account or account information.
        \item \textsf{Circumventing and abusing moderation mechanism}: Misusing moderation tools to submit false reports, undermining their intended purpose, or engaging in any actions designed to circumvent imposed penalties or disciplinary measures, such as creating new accounts to avoid restrictions.
        \item \textsf{Fraud and scamming}: Intentional deception or misrepresentation that unfairly acquires assets, information, or any form of advantage by misleading others, including phishing, social engineering, etc.
        \item \textsf{Law violation}: Violation of any local, state, national, or international laws or regulations.
        \item \textsf{Unauthorized transaction and commercialization}: Transactions and commercialization without the service provider's authorization, such as buying, selling, trading, sharing, or transferring account access, real money or virtual currency transactions, and cross-server trades.

    \end{squishitemize}
    \item \textsf{Moderation}: The procedure and mechanism for reviewing and monitoring player behaviors as well as the consequence if rules of the game, its community and service provider are violated.
      \begin{squishitemize}
    \item \textsf{Moderation Consequence}: The actions and penalties imposed when rules are violated
    \item \textsf{Moderation Mechanism}: The tools, processes, and systems used to enforce rules
  \end{squishitemize}
    
    \item \textsf{Expected conduct}: The accepted and expected behaviors which are aligned with the norms and rules of the game, its community and service provider.
    \item \textsf{Value justification}: The justification provided for a safety rule, particularly the values and core beliefs of the game, its community and service provider
\end{squishitemize}

\subsection{Inter-Rater Reliability in Reference Dataset}
~\label{sec_appendix_irr}
\begin{table}[h]
    \footnotesize
    \setlength{\tabcolsep}{3.5pt}
    \renewcommand{\arraystretch}{0.9}
    \centering
    \caption{ Inter-rater reliability (Cohen’s $\kappa$) and label distribution in \workname's reference dataset (pre-curation).}
    \label{tab:label_stats}
    \begin{tabular}{m{4.3cm} >{\centering\arraybackslash}m{1.2cm} 
                    >{\centering\arraybackslash}m{1.5cm}}
    \toprule
        \textbf{Label} & \textbf{$\kappa$} & \textbf{Count} \\
    \midrule

    \rowcolor{blue!35}\multicolumn{3}{l}{\textbf{Misconduct}} \\
    \rowcolor{blue!10}\quad Misconduct (Topic) & 0.673 & 508 \\
    \quad Harassment and threat & 0.620 & 116 \\
    \quad Hate and discrimination & 0.517 & 46 \\
    \quad Exploiting and cheating & 0.768 & 117 \\
    \quad Abuse of play and antagonistic play & 0.466 & 47 \\
    \quad Circumventing and abusing moderation mechanism & 0.685 & 34 \\
    \quad Inappropriate content creation and sharing & 0.705 & 186 \\
    \quad Privacy breach & 0.691 & 51 \\
    \quad Impersonation and identity theft & 0.660 & 60 \\
    \quad Unauthorized transaction and commercialization & 0.674 & 43 \\
    \quad Fraud and scamming & 0.812 & 38 \\
    \quad Law violation & 0.779 & 63 \\

    \rowcolor{green!60}\multicolumn{3}{l}{\textbf{Moderation}} \\
    \rowcolor{green!20}\quad Moderation (Topic) & 0.698 & 305 \\
    \quad Moderation consequence & 0.820 & 168 \\
    \quad Moderation mechanism   & 0.752 & 111 \\

    \rowcolor{gray!20}\multicolumn{3}{l}{\textbf{Positive governance}} \\
    \quad Values justification & 0.612 & 174 \\
    \quad Expected conduct & 0.766 & 249 \\

    \bottomrule
    \end{tabular}
\end{table}

\subsection{Template for NLI Classification}
\label{sec_appendix_nli_template}

\begin{tcolorbox}[beamerblockstyle, colframe=orange!30, colback=orange!30]
\footnotesize
{
\textbf{\texttt{Input:}} 
\begin{squishitemize}
    \item \textbf{\texttt{premise:}} Exploitation of any new or known bug or glitch for personal gain is strictly forbidden and may result in character rollback, Account suspension or revocation. 
    \item \textbf{\texttt{hypothesis:}} The text is about Misconduct
\end{squishitemize}
\textbf{\texttt{Output:}} yes
}
\end{tcolorbox}

\subsection{Classification Model Training Configuration}
~\label{sec_appendix_model_config}
To enable parameter-efficient adaptation for training \workname's NLI classifier, 
our implementation relies on \textsc{LoRA} to restrict trainable parameters, with rank $r=8$ and $\alpha=16$ on the query and value projection matrices, while keeping all other parameters frozen.
Fine-tuning is performed with a learning rate of $3\times10^{-4}$ and a batch size of 8. 
To maximize data utility and avoid overfitting, we adopt LabelWave \cite{yuan2025early}, a validation-free early stopping method that monitors prediction changes within the training set. 
As a result, it takes 38 epochs in our modeling fine-tuning.

\subsection{Template for QA Query}
\label{sec_appendix_qa_template}

\begin{tcolorbox}[beamerblockstyle, colframe=orange!30, colback=orange!30]
\footnotesize
{
\textbf{\texttt{Input:}} 
\begin{squishitemize}
    \item \textbf{\texttt{Context:}} Exploitation of any new or known bug or glitch for personal gain is strictly forbidden and may result in character rollback, Account suspension or revocation. 
    \item \textbf{\texttt{Query:}} What technical or gameplay features and tools are exploited in the misconduct? 
\end{squishitemize}
\textbf{\texttt{Output:}} bug, glitch
}
\end{tcolorbox}

\subsection{Coverage of Moderation Strategies}

\begin{figure}[H]
    \centering
    \includegraphics[width=\columnwidth]{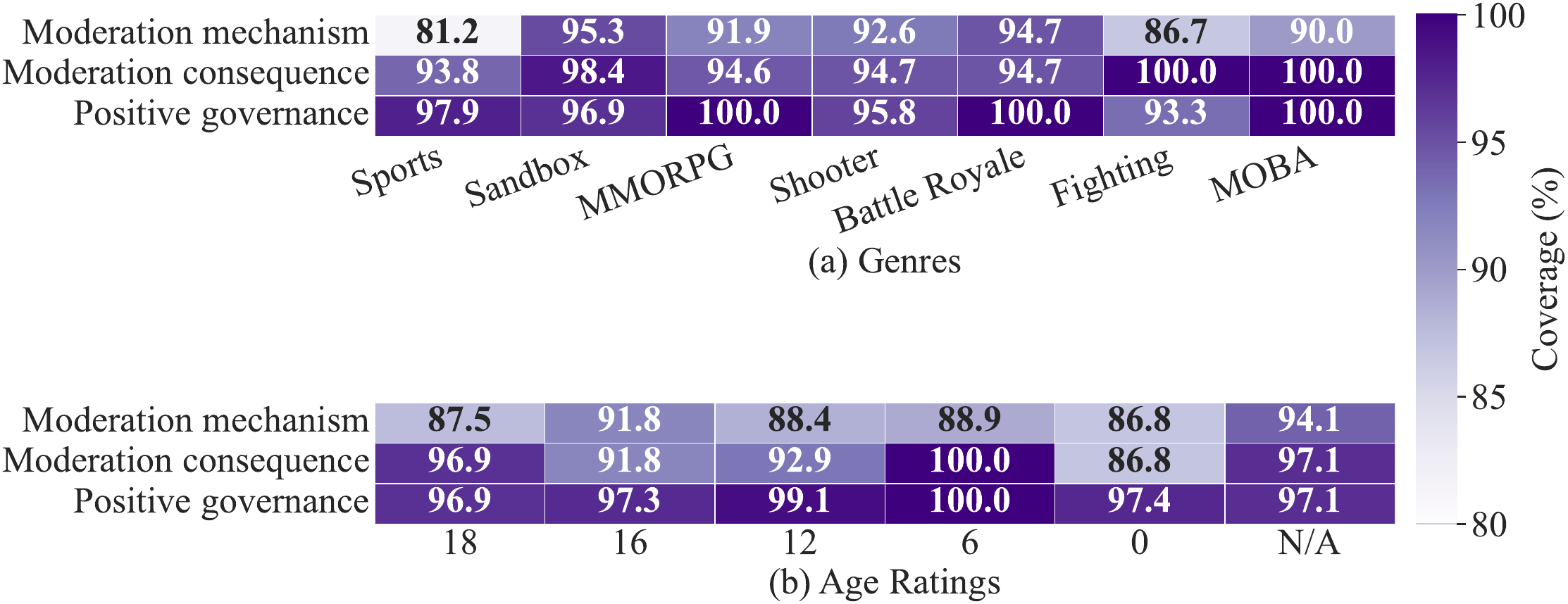}
    \caption{Coverage of moderation labels across game attributes. Each panel shows per-game coverage (\%) by (a) genres and (b) age ratings, with darker cells indicating higher prevalence of the corresponding moderation label.} 
    \label{fig:Moderation_wide}
\end{figure}

\subsection{Search Terms}
\label{keyword_list_lexicon}

We construct a lexicon of root forms for toxicity-related framing developed in prior work~\cite{wijkstra_rogers_mandryk_veltkamp_frommel_2024} and perform root-based matching to capture toxicity-related complaints in player review that matches any of their morphological variants (e.g., toxicity and harassing): \texttt{toxic*, harass*, insult*, grief*, trol*, 
offen*, inappropriate, abus*, flam*}

Similarly, our search terms for child-related CoC content are: 
\texttt{child*, kid*, minor*, youth*, youngster*, juvenile*, underage*, under 18, under-age*, young player*, young user*"}

%


\end{document}